\documentclass[10pt]{article}

\usepackage{my-common}

\newcommand{\blind}{1}

\begin{document}

\def\spacingset#1{\renewcommand{\baselinestretch}%
{#1}\small\normalsize} \spacingset{1}

\if1\blind
{
  \title{\bf Bayesian Nonparametric Multivariate Spatial Mixture Mixed Effects Models
    with Application to American Community Survey Special Tabulations}
  \author{Ryan Janicki$^{a}$, Andrew~M. Raim$^{a}$, Scott H. Holan$^{bc}$, and Jerry
    Maples$^{a}$
    \vspace{0.5em} \\
    $^{a}$Center for Statistical Research and Methodology, U.S. Census Bureau \\
    $^{b}$Department of Statistics, University of Missouri\\
    $^{c}$Office of the Associate Director for Research and Methodology, U.S. Census
    Bureau 
  }
  \maketitle
}\fi

\if0\blind
{
  \bigskip
  \bigskip
  \bigskip
  \begin{center}
    {\LARGE\bf Bayesian Nonparametric Multivariate Spatial Mixture Mixed Effects
      Models with Application to American Community Survey Special Tabulations}
  \end{center}
  \medskip
}\fi

\bigskip
\begin{abstract} 
  Leveraging multivariate spatial dependence to improve the precision of estimates using
  American Community Survey data and other sample survey data has been a topic of recent interest
  among data-users and federal statistical agencies. One strategy is to use a multivariate
  spatial mixed effects model with a Gaussian observation model and latent Gaussian process
  model. In practice, this works well for a wide range of tabulations. Nevertheless, in
  situations that exhibit heterogeneity among geographies and/or sparsity in the data, the
  Gaussian assumptions may be problematic and lead to underperformance. To remedy these
  situations, we propose a multivariate hierarchical Bayesian nonparametric mixed effects
  spatial mixture model to increase model flexibility. The number of clusters is chosen
  automatically in a data-driven manner. The effectiveness of our approach is demonstrated
  through a simulation study and motivating application of special tabulations for American
  Community Survey data.
\end{abstract}

\noindent%
 \vfill

\newpage

\spacingset{1.5}
\section{Introduction}
\label{sec:intro}

The American Community Survey (ACS) is the largest household survey run by the U.S. Census Bureau. The ACS is an ongoing survey which samples approximately 3.5 million households
annually spread-out through the year and collects data on a broad range of social, demographic,
economic, and housing characteristics.%
\footnote{For more details see \url{http://www.census.gov/acs}.}  The ACS annually produces a
large number of tables at various levels of aggregation. Specifically, the ACS produces both 1-year
and 5-year period estimates depending on the population for different geographies. The one-year
period estimate is derived over a single calendar year for geographical areas with a population
of at least 65,000.  In contrast, for all geographical areas down to the tract level, table
estimates are also produced aggregating 5 years of survey data (also known as 5-year period
estimates); e.g., the 2018 5-year ACS estimates will be tabulated using respondents from
January 2014 to December 2018. The Census Bureau also produced 3-year period estimates for
geographies with populations between 20,000 and 65,000, though these estimates were
discontinued in 2013
(\url{https://www.census.gov/programs-surveys/acs/guidance/estimates.html}).

The U.S. Census Bureau publishes tables using 5-year ACS estimates for many social,
demographic, and economic cross-classifications, when the sample sizes are sufficiently large
for the published estimates to be considered reliable, and to not pose a risk of disclosure of
a respondent's personal identifying information.  In addition to the standard tables released
by the U.S. Census Bureau, stakeholders have requested custom statistical data products known
as {\it special tabulations}, which are more detailed than publicly available tables. For
example, the Minnesota Department of Education requested a tabulation to help administer
programs for early childhood development.  This tabulation is associated with a
finer breakdown of children's ages into age groups (0--1, 2--3, 4--5) crossed by race, income
level, Hispanic origin, or relationship to householder (individually, not jointly) for every
county in Minnesota.  Many of these tables have cells with very few or no survey cases,
resulting in issues of data quality and/or disclosure limitations.  From a data quality
perspective, estimates which only use the direct survey data based on a small sample size may
not be precise or may not be possible at all \citep{rao15}.  Releasing estimates based on small
sample sizes also increases the risk of unintended disclosures for responding individuals.

The mission of the Census Bureau is to provide accurate official statistics using data
collected by its programs, but also to ensure that privacy and confidentiality of respondents
is protected. For this reason, the release of special tabulations products was significantly
curtailed, and the agency is considering alternatives to releasing the direct survey estimates.
The Census Bureau is beginning to incorporate techniques from the
differential privacy literature which offer mathematical guarantees on privacy
\citep{abowd2018us}. There has been a large-scale and ongoing effort to develop innovative
methods to protect releases of major Census Bureau data products.  This work considers a more
immediate solution, which is to produce model-based predictions, based on the direct estimates,
for release by the agency. Because model-based predictions are indirect estimates which utilize
the entire dataset, the individual disclosure risk is greatly reduced, albeit without the
mathematical guarantees of differential privacy.



Latent Gaussian process (LGP) models have become a standard tool for modeling dependencies in
count-valued and other non-Gaussian datasets, see \citet{diggle1998model}. A natural approach
to implementing LGP models is through hierarchical statistical modeling and, in particular,
using a Bayesian formulation.  In this context, the joint distribution of the data, latent
processes, and unknown parameters are written as the product of a data model, a Gaussian
process model, and a parameter model (e.g., see \citet{cressie2011statistics};
\citet{banerjee2014hierarchical}, among others). Within this model framework, complex
dependencies based on the multivariate structure of the outcome, spatial and temporal
relationships, and their interactions, can be incorporated. Efficient estimation of the joint
posterior distribution of the process and parameters given the data then proceeds through an
application of Bayes theorem.

Many models used in small area estimation (SAE) exist within this class of LGP models, e.g. the
Fay-Herriot model \citep{fay79} and its extensions; see \citet{BradleyEtAl2015} for additional
discussion. The LGP class of models is highly flexible. They can include many types of
relationships in the data and provide a useful tool for `borrowing strength,' using
multivariate, spatial and/or temporal dependencies, to improve estimates that lack sufficient
survey data.

Another complexity of the data is that some tabulations may not result from a singular spatial
pattern, but instead may result from the aggregate of several. In other words, modeling the
distribution of tabulated values may not conform to an underlying known parametric
distribution. For example, preliminary analysis of the age by race table suggested that there were different spatial patterns for different racial groups across the counties. To address this data
complexity, we propose a Bayesian nonparametric mixture of LGP
models. Sections~\ref{sec:msmmix} and \ref{sec:s48405} will explore this further.


Related work is that of \citet{gel05}, who introduced Dirichlet process mixing for spatial
point process models.  \citet{gel05} used the Dirichlet process prior for the purpose of
developing a framework for the analysis of non-Gaussian, nonstationary, point-referenced
spatial data.  In contrast, the current work is concerned with analysis of multivariate spatial
areal data collected from a sample survey.  We combine elements of small area estimation theory
\citep{rao15}, multivariate spatial distribution theory \citep{BradleyEtAl2015}, and Bayesian
nonparametrics \citep{hjort2010bayesian} for the purpose of producing flexible model-based
predictions of area-level means with greater precision than those of the direct, survey-based
estimates.  We introduce a Dirichlet process prior on the latent Gaussian process, primarily
for the purpose of clustering the observed data on multivariate characteristics and on similar
spatial patterns.

The remainder of this paper is organized as follows.  In Section~\ref{sec:MSM}, we describe
multivariate spatial mixed effects models (MSM) and their application to the ACS.  In
Section~\ref{sec:msm-st}, we present results of fitting the MSM to two different 5-year ACS
special tabulations.  In Section~\ref{sec:s48400} we fit the MSM to ACS 5-year estimates of the
number of children by age in counties in Minnesota, and show good performance of model-based
estimates, compared the corresponding direct, survey-based estimates.  In
Section~\ref{sec:msmfail}, we show that for certain special tabulations, the MSM can produce
predictions of obvious poor quality when the data exhibit heterogeneous spatial and
multivariate patterns.  As an example, the MSM is fit to ACS 5-year estimates of the number of
children in counties by age and race in counties in Minnesota.  To remedy such problems, an
extension to the MSM is proposed in Section~\ref{sec:msmmix}, which introduces the multivariate
spatial mixed effects model with Dirichlet process mixing (MSMM).  An empirical simulation
study is provided in Section~\ref{sec:empirical} and illustrates the effectiveness of our
proposed modeling approach.  In Section~\ref{sec:s48405}, we fit the MSMM to the age by race
dataset, and show good performance of the predictions, compared to the predictions from the
MSM or the direct estimates.  Concluding remarks are given in Section~\ref{sec:conc}.

\section{Multivariate Spatial Mixed Effects Model}\label{sec:MSM}

Conceptually, special tabulations and other ACS data are a collection of multi-way contingency
tables for a set of areas in a given geographical domain. For example, one particular
tabulation concerns the counts of children in three groups: 0--1, 2--3, and 4--5 years of
age. A one-way table with counts of the three age groups is constructed for each county in the
United States. A second tabulation provides counts of children in the same three age groups,
but also cross-classified by a race factor with seven categories: White alone, Black alone,
Asian alone, American Indian or Alaska Native alone, Native Hawaiian or Pacific Islander alone,
Other alone, or two or more races.  Here, a two-way table is constructed for each county. These
are just two examples of many special tabulations of ACS data that the U.S. Census Bureau is
tasked with producing; others include cross-classifications of demographic characteristics
(e.g.~age, race, gender, income), different housing characteristics (e.g.~owner vs.~renter) and
geographic regions (e.g.~states, counties, and tracts).

In general, suppose there are $k$ factors in a particular table under consideration, where the
levels of the $j$th factor are indexed by $i_j = 1, \ldots, I_j$. Excluding marginal counts,
which are typically provided with the data, let $L$ denote the number of interior table
cells. In this work, factors are crossed so that $L = I_1 \times \cdots \times I_k$; however,
in principle, factors may also be nested so that certain levels of one factor are defined only
for some levels of other factors. Let $\mathcal{D}$ denote the collection of geographies
currently under consideration, writing $A \in \mathcal{D}$ to represent a particular areal unit
within the domain. Each table consists of direct estimates based on the Horvitz-Thompson
estimator
\begin{equation}\label{E:direct}
  Z^{*(l)}(A) = \sum_{j \in \mathcal{S}^{(l)}(A)} w_j,
\end{equation}
where \(j \in \mathcal{S}^{(l)}(A)\) are the sampled units belonging to interior table cells $l
= 1, \ldots, L$ and areas $A \in \mathcal{D}$, and \(w_j\) is the associated survey weight.
Let \(D^{*(l)}(A)\) be an estimate of the design-based variance of \(Z^{*(l)}\); the
U.S. Census Bureau uses the successive differences replication method \citep{jud90, fay95,
  tor14}.  Therefore, the multivariate observation $\{ \left( Z^{*(l)}(A), D^{*(l)}(A) \right)
: l = 1, \ldots, L \}$ represents a table of direct estimates for area $A$. Let \(m = \left|
\mathcal{D} \right|\) denote the number of areal units so that \(n = mL\) is the total number
of observations in the tabulation. To facilitate modeling estimates of counts, which are likely
to be right-skewed, we transform the direct estimates \eqref{E:direct} using
\begin{equation}\label{E:logDirect}
  Z^{(l)}(A) = \log \left( Z^{*(l)}(A) + 1 \right),
\end{equation}
adding one due to the presence of direct estimates with a value of zero.  Let \(D^{(l)}(A)\) be
the variance estimate of the log-transformed direct estimate, \(Z^{(l)}(A)\).

To produce model-based estimates of the direct ACS estimates, we must formulate an appropriate
model. Because the data represent entries of spatially-dependent contingency tables, we
consider a multivariate spatial mixed effects model (MSM) which takes these dependencies into
account. The model will now be formulated via the standard data, process, parameter model
formulation popularized in the spatial statistics literature \citep{cressie2011statistics}. The
data model is defined as
\begin{equation}\label{E:data}
  Z^{(l)} (A) = Y^{(l)} (A) + \varepsilon^{(l)} (A),
\end{equation}
for \(A \in \mathcal{D}\) and \(l = 1, \dots, L\). Survey estimates $Z^{(l)}(A)$ are assumed to
be design unbiased estimates of the underlying population quantities $Y^{(l)}(A)$. The survey
design is incorporated via sampling error terms, \(\varepsilon^{(l)} (A)\), which are assumed
to be independent, normally distributed, mean-zero random variables, with known sampling
variances \(D^{(l)} (A)\). The process model is given by
\begin{equation}\label{E:process}
  Y^{(l)} (A) = \mu^{(l)} (A) + \nu^{(l)} (A),
\end{equation}
where \(\mu^{(l)} (A)\) is an unknown fixed effect representing the large-scale multivariate
spatial trend. The term \(\nu^{(l)} (A)\) is a random effect representing the fine-scale
variability, and is used to incorporate multivariate spatial dependencies in the process model.
The fixed effects, \(\mu^{(l)} (A)\), are modeled using a linear regression, \(\mu^{(l)}(A) =
\boldsymbol{x}^{(l)} (A)^\top \boldsymbol{\beta}\), where \(\boldsymbol{x}^{(l)}(A)\) is a
\(p\)-dimensional vector of known covariates, and \(\boldsymbol{\beta}\) is a \(p\)-dimensional
vector of unknown regression coefficients. Prediction of the latent process, \(Y^{(l)}(A)\),
from the observed data, \(Z^{(l)} (A)\), is the primary interest in fitting the model.  After
model fitting, the inverse transformation
\begin{equation}\label{E:inverse}
  Y^{*(l)}(A) = e^{Y^{(l)}(A)} - 1
\end{equation}
can be applied to obtain predictions and measures of uncertainty on the original scale.

\begin{remark}
  In the spatial literature, but not necessarily in the small area estimation literature, it is
  common to model the random process, \(Y^{(l)} (A)\), using an additional fine-scale
  variability term, \(\xi^{(l)} (A)\), so that the process model becomes
  \begin{equation*}
    Y^{(l)} (A) = \mu^{(l)} (A) + \nu^{(l)} (A) + \xi^{(l)} (A),
  \end{equation*}
  where, in many cases, \(\xi^{(l)} (A) \iid \text{N}(0, \sigma^2_\xi)\).  The term \(\xi^{(l)}
  (A)\) is used to describe the local behavior of the process \(Y^{(l)} (A)\).  In the work
  presented here, fitting a spatial model that includes the fine-scale variability terms for
  the ACS data resulted in overfitting, so that predictions and uncertainty estimates were
  largely the same as the survey-based estimates.  It is possible that the sampling variances
  in the data model largely account for any residual fine-scale variability, so that inclusion
  of the additional terms, \(\xi^{(l)} (A)\), result in a weakly identifiable model. In
  addition, at the level of geographies often considered, when constructing tabulations for
  public dissemination (county-level), the spatial variation is generally smooth and can be
  accounted for through the other terms in the model.  When using the process model in
  \eqref{E:process}, we found no issues with overfitting.  \xqed{$\blacksquare$}
\end{remark}

The random effects, \(\nu^{(l)}(A)\), are modeled using a basis expansion,
\begin{equation*}
  \nu^{(l)}(A) = \boldsymbol{\psi}^{(l)}(A)^\top \boldsymbol{\eta},
\end{equation*}
where the \(\boldsymbol{\psi}^{(l)}(A)\) are \(r\)-dimensional multivariate spatial basis
functions and the distribution of the random effect \(\boldsymbol{\eta}\) is specified to
capture spatial dependencies in the data.  Taking \(r \ll n\) has the effect of inducing
sparsity and reducing the rank of the model, which leads to a more parsimonious model,
drastically reducing the computational burden of fitting the model, particularly when fitting
to very large datasets \citep{hug13}.  Also, when \(r \ll n\), multiple observations will share
common realizations of the random effects in this model specification, inducing multivariate
dependence.

The vectors \(\boldsymbol{\psi}^{(l)} (A)\) can be any set of multivariate spatial basis
functions \citep[see][]{bradley2017regionalization}.  However, following \citet{hug13} and
\citet{BradleyEtAl2015}, we use the Moran's I basis functions, which are chosen to avoid
confounding between the fixed effects and the random effects.  Let \(\boldsymbol{A}\) be an \(n
\times n\) multivariate adjacency matrix corresponding to areal units \(A \in \mathcal{D}\);
specifically, \(\vec{A} = \vec{W} \otimes \vec{1}_L \vec{1}_L^\top\) where $\otimes$ represents
the Kronecker product, $\vec{1}_L$ is an $L \times 1$ matrix of ones, and $\vec{W} = (w_{ii'})$
is a standard adjacency matrix defined by
\begin{align*}
  w_{ii'} =
  \begin{cases}
    0 & \text{if $i = i'$}, \\
    1 & \text{if $i \neq i'$ and areas $i$ and $i'$ are adjacent}, \\
    0 & \text{otherwise},
\end{cases}
\end{align*}
for $i, i' \in \{ 1, \ldots, m \}$. The Moran's I (MI) operator \citep{mor50} is
given by
\begin{equation}\label{E:moranI}
  G(\boldsymbol{X}, \boldsymbol{A}) \equiv (\boldsymbol{I}_n - \boldsymbol{X}
    (\boldsymbol{X}^\top \boldsymbol{X})^{-1} \boldsymbol{X}^\top) \boldsymbol{A}
    (\boldsymbol{I}_n - \boldsymbol{X} (\boldsymbol{X}^\top \boldsymbol{X})^{-1}
    \boldsymbol{X}^\top). 
\end{equation}
The multivariate spatial basis functions \(\boldsymbol{\psi}^{(l)} (A)\) are defined to be the
\(r\) eigenvectors corresponding to the \(r\) largest positive eigenvalues of
\eqref{E:moranI} \citep{hug13}.

We assume \(\boldsymbol{\eta} \sim \text{N}_r (\boldsymbol{0}, \sigma^2_{\boldsymbol{\eta}}
\boldsymbol{K})\), where \(\boldsymbol{K}\) is a positive definite matrix chosen to induce a
conditional autoregressive structure on the random effects, \(\nu^{(l)}(A)\).  The variance
component, \(\sigma^2_{\boldsymbol{\eta}}\), is an unknown parameter to be estimated.  Let
\(\boldsymbol{\Psi}\) be the \(n \times r\) matrix with rows consisting of the multivariate
spatial basis functions, \(\boldsymbol{\psi}^{(l)}(A)\), and let \(\boldsymbol{Q}\) be the
singular, positive semi-definite precision matrix for an intrinsic conditional autoregressive
(ICAR) process.  That is, \(\boldsymbol{Q} = \boldsymbol{D} - \boldsymbol{A}\), where
\(\boldsymbol{D}\) is a diagonal matrix, with diagonal entries equal to the row sums of
\(\boldsymbol{A}\).  Following \citet{hug13}, let \(\boldsymbol{K}^{-1} =
\boldsymbol{\Psi}^\top \boldsymbol{Q} \boldsymbol{\Psi}\).  It can be shown that
\(\boldsymbol{K}^{-1}\) is positive definite, so long as \(\boldsymbol{X}\) includes an
intercept \citep{por15}.  \citet{BradleyEtAl2015} shows that this specification of the
precision matrix \(\boldsymbol{K}^{-1}\) minimizes the Frobenious norm \(\lVert \vec{Q} -
\boldsymbol{\Psi} \vec{C}^{-1} \boldsymbol{\Psi}^\top \rVert_\text{F}\) over $\vec{C}$ in the
space of positive definite \(r \times r\) matrices. In this sense, \(\boldsymbol{K}^{-1}\) is
the best positive approximant to the ICAR precision matrix \(\boldsymbol{Q}\)
\citep{Higham1988}.

The model specification is completed by choosing parameter models for the unknown parameters,
\(\boldsymbol{\beta}\) and \(\sigma^2_{\boldsymbol{\eta}}\).  Using independent priors,
\(\boldsymbol{\beta} \sim \text{N}_p(\boldsymbol{0}, \sigma^2_{\boldsymbol{\beta}}
\boldsymbol{I}_{p \times p})\), and \( \sigma^2_{\boldsymbol{\eta}} \sim \text{IG}(a, b)\),
leads to full conditional distributions from known parametric families, allowing for easy
sampling from the posterior distribution.  See the Supplementary Materials for derivation of
the full conditional distributions.

\section{Application of MSM to ACS Special Tabulations}
\label{sec:msm-st}

The MSM model from Section~\ref{sec:MSM} can be a useful tool for producing model-based special
tabulations of ACS data. Model-based predictions may be significantly more accurate than
associated ACS direct estimates for certain demographic and geographic cross-classifications of
ACS data. However, MSM predictions for some tabulations can be of extremely poor quality. In
this section, we consider MSM for two particular tabulations. Section~\ref{sec:s48400} models
total number of children by age group. Section~\ref{sec:msmfail} models total number of
children by the cross-classification of both age group and race. Both tabulations are for
counties in Minnesota, and are based on 2015 ACS 5-year data. The 2015 5-year ACS data consists
of pooled ACS data over the period 2011--2015, with survey weight adjustments made to reflect
the different time periods in which data was collected.  The advantage of using 5-year data
over 1-year ACS data is the larger sample sizes available, and that 5-year data is publicly
available for a wider range of geographies and economic and demographic variables than 1-year
data.

We find that MSM can produce high quality predictions in the situation of
Section~\ref{sec:s48400}, with model-based predictions having greatly reduced standard errors,
relative to the direct estimates.  We have also found (but not included in the paper) that MSM
can produce high quality predictions when applied to certain cross-classifications such as age
by poverty status, age by gender, and age by housing composition using similar
datasets. However, poor predictions from the fitted MSM model could occur when the dimension of
the special tabulations increased, when the number of cross-classifications of interest
increased, or when---even with relatively simple multivariate tabulations---the underlying
spatial field assumed by the model was inappropriate. This situation is illustrated in
Section~\ref{sec:msmfail}.

\subsection{Estimation of the number of children in counties in
  Minnesota}\label{sec:s48400}

We first consider fitting a special tabulation of one-way contingency tables containing counts
of children in categories 0--1, 2--3, and 4--5 years of age for counties in Minnesota.  Within
counties, there is strong positive correlation of direct estimates of total children in the
three age categories.  Exploratory analysis using Moran's I statistic \citep{mor50} on each
marginal dataset also indicated strong spatial correlation in the data.

\begin{figure}[t]
  \centering
  \begin{subfigure}{0.45\textwidth}
    \centering
    \includegraphics[width=\textwidth, height=2in]{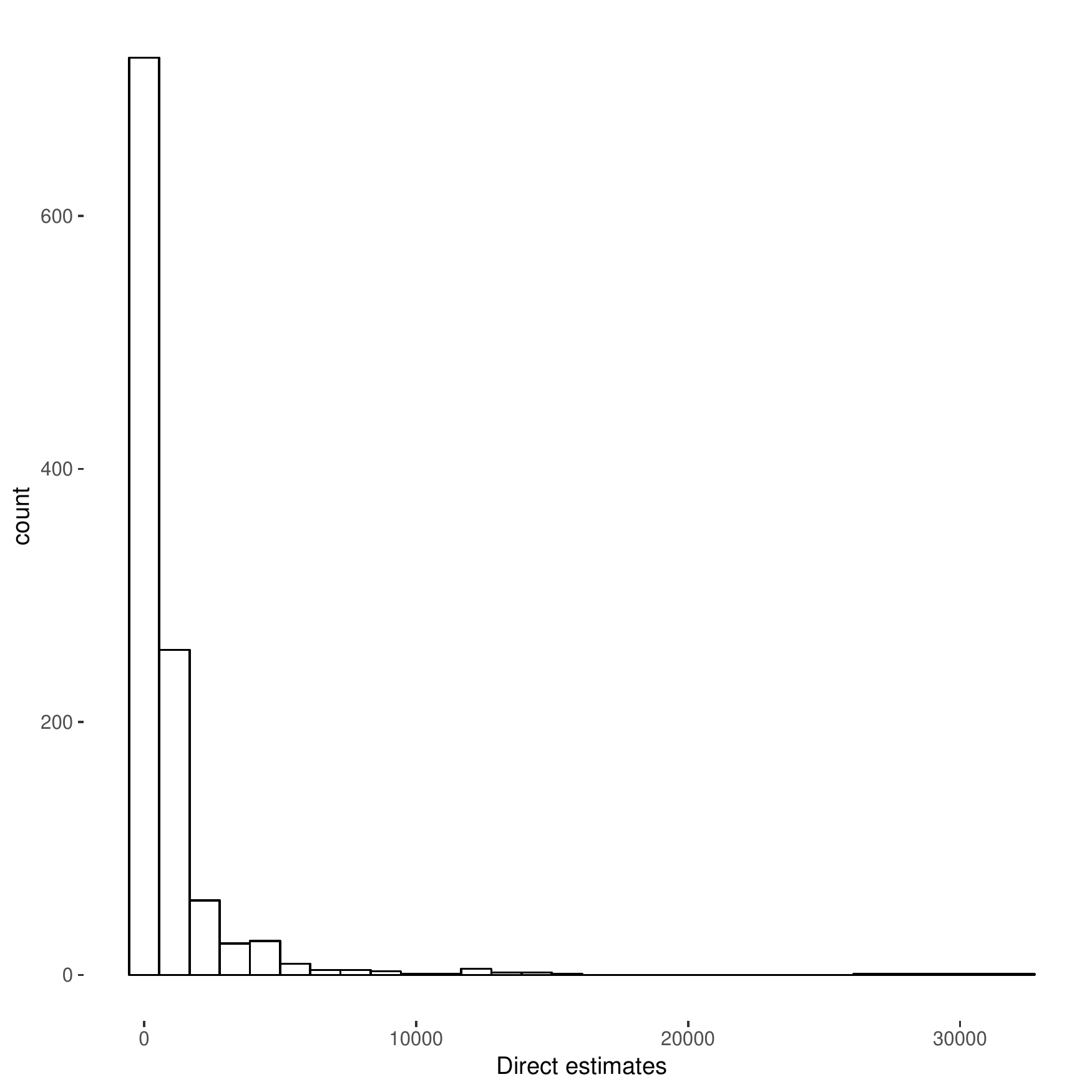}
    \caption{Histogram of the ACS 5 year estimates of total children, ages 0--1, 
      2--3, and 4--5, in counties in Midwestern states}\label{fig:directHist}
  \end{subfigure}
  ~
  \begin{subfigure}{0.45\textwidth}
    \centering
    \includegraphics[width=\textwidth, height=2in]{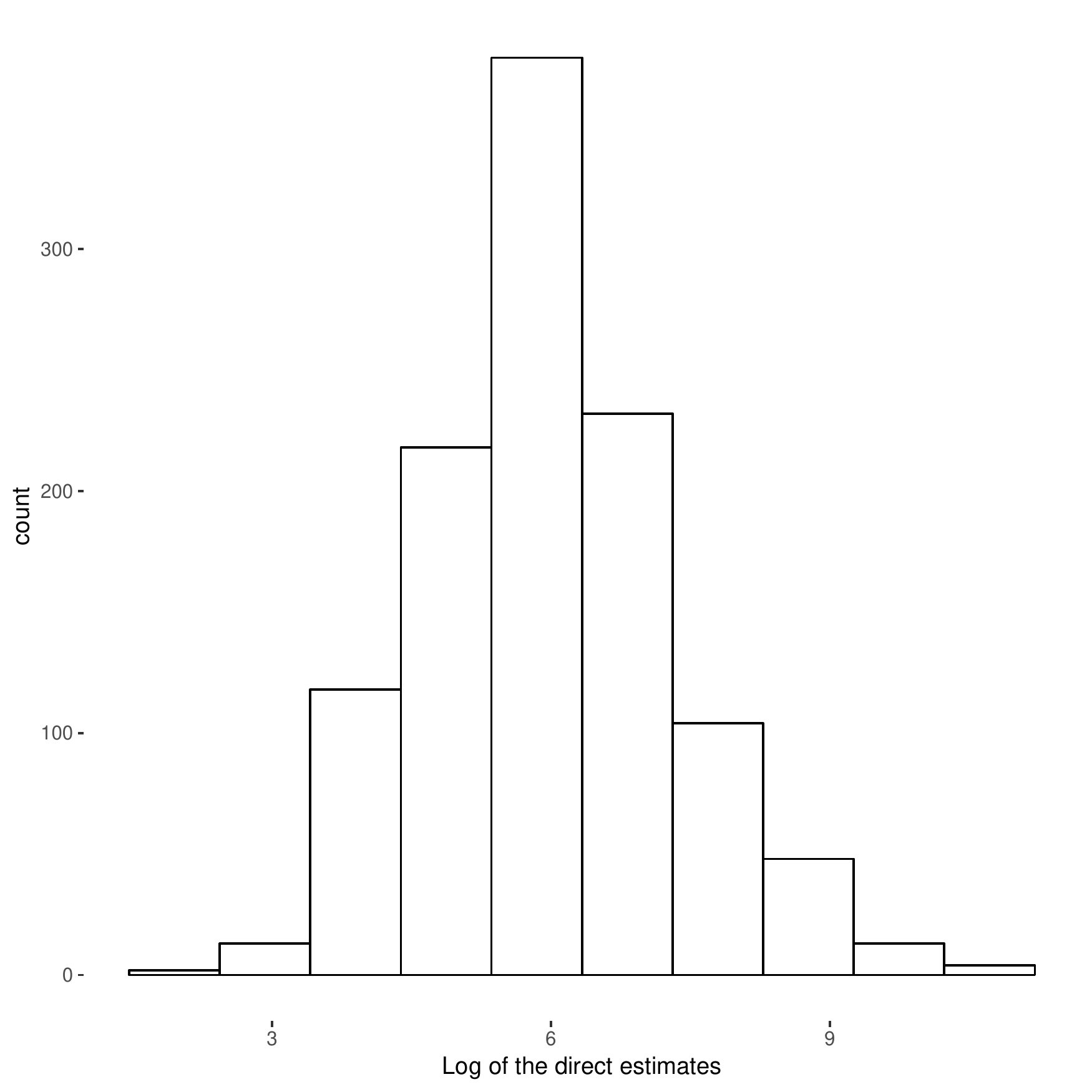}
    \caption{Histogram of the log of the ACS 5 year estimates of total children, ages
      0--1, 2--3, and 4--5, in counties in Midwestern
      states}\label{fig:logDirectHist}
  \end{subfigure}
  \caption{Comparison of the distribution of estimates of counts with the
    distribution of the estimates of the log of counts.}
  \label{fig:hist}
\end{figure}

Figure~\ref{fig:directHist} shows a histogram of the direct estimates of total children in
counties in the 0--1, 2--3, and 4--5 age groups. Before applying transformation
\eqref{E:logDirect} the distribution of the direct estimates is heavily right skewed due to the
presence of counties with large cities.  Figure~\ref{fig:logDirectHist} shows the histogram of
the transformed direct estimates; this suggests modeling the direct estimates on the log scale
using the multivariate spatial model.

\begin{figure}[t]
  \centering
  \includegraphics[height=3in, width=0.65\textwidth]{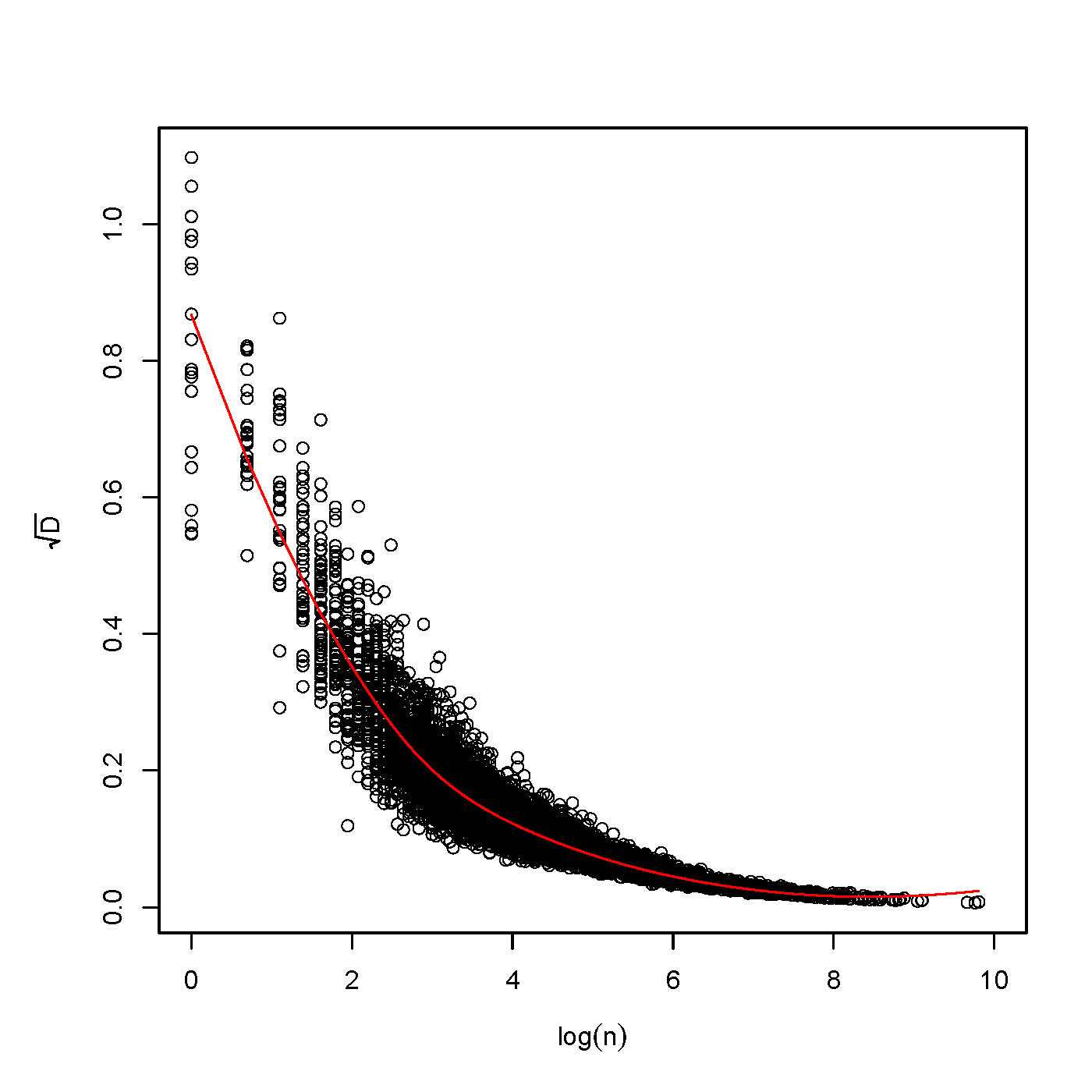}
  \caption{Scatter plot of the direct estimates of the log counts of children, ages 0--1, 2--3, or 4--5, in counties in Minnesota and surrounding states, vs. the log of the
    sample sizes.  The red line shows the fit of a LOESS regression of the direct estimates of
    the log counts on the log of the county sample sizes.}
  \label{fig:genVar}
\end{figure}

The sampling variances of the direct estimates are also needed as model inputs.  For the
nonzero direct estimates, the method of replicate weights can be used to estimate the sampling
variances \citep{jud90}.  However, the presence of direct estimates of zero introduces an
additional challenge, due to the fact that there is no way to directly estimate their sampling
variances.  To overcome this difficulty, a generalized variance function is used.
Figure~\ref{fig:genVar} shows a scatter plot of the estimated sampling variances versus the log
of the sample sizes for the nonzero direct estimates. For the areas where the estimated
sampling variances are not defined, a plug-in estimate from the displayed LOESS smoothing curve
is used in the model.

While the goal is estimation of counts in counties in Minnesota, we fit the model to data from
counties in Minnesota, as well as counties in the surrounding states of Wisconsin, Iowa, North
Dakota, and South Dakota.  Expanding the set of areal units to the counties in surrounding
states resulted in greater precision of predicted values, as the spatial field seemed to be
more easily identified with an expanded number of spatial regions.  We also investigated
expanding the number of included areal units to all counties in the continental United States.
However, this drastically increased the computational time needed to fit the model, without
noticeably changing the predictions.  We found expanding the set of areal units to include only
counties in adjacent states to be a good tradeoff, in terms of reducing computational
burden while maintaining stability of predictions.

The covariate, \(\boldsymbol{x}^{(l)}(A)\), used in the model includes an intercept, the log of
the county total population for area \(A\), which is assumed known from Census population
estimates, and indicator variables encoding the combination of factors for entry \(l\) of the
original contingency table.  In this case, \(l = 1, 2, 3\) corresponds to age groups 0--1,
2--3, and 4--5.

\begin{figure}[t]
  \centering
  \begin{subfigure}{0.45\textwidth}
    \centering
    \includegraphics[width=\textwidth, height=2in]{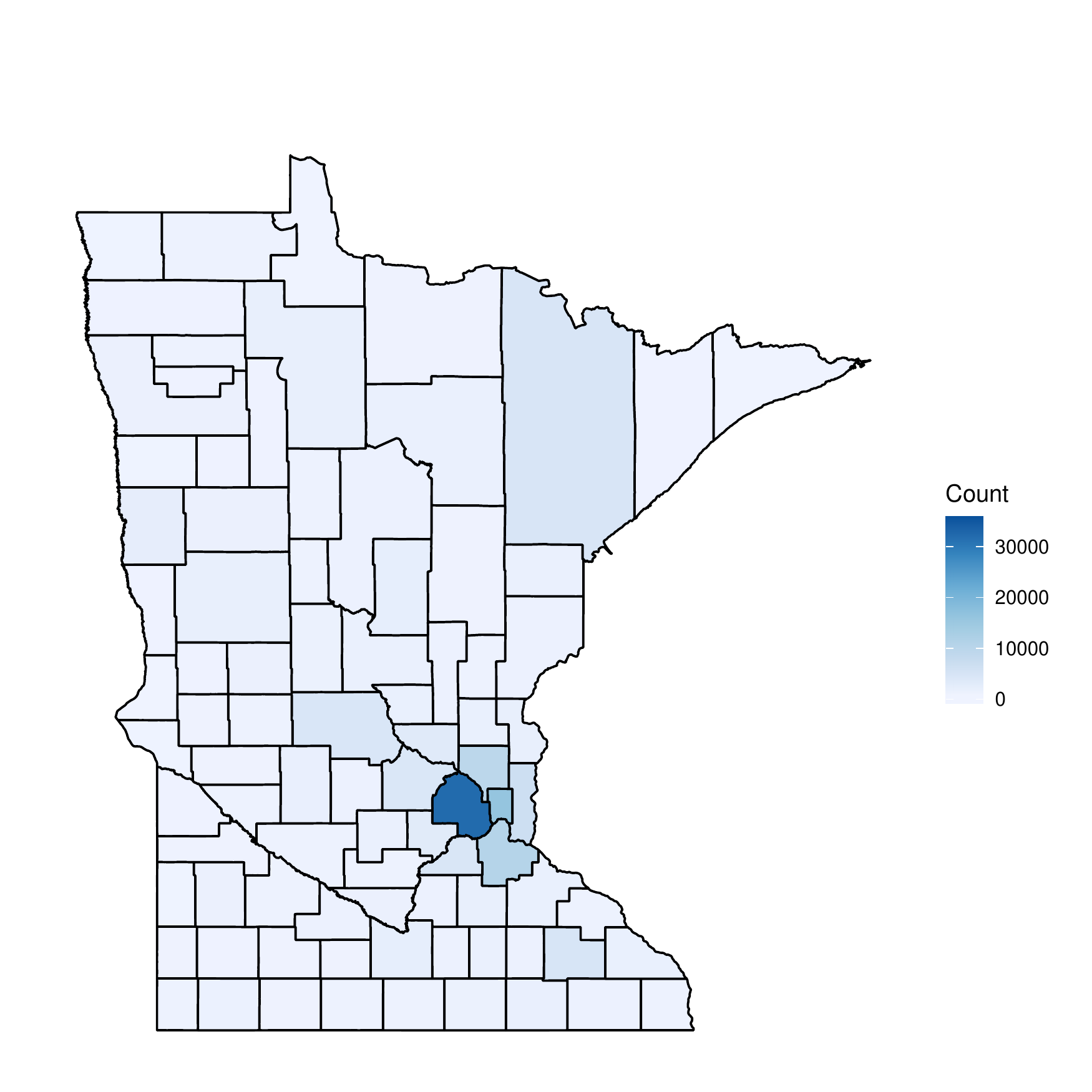}
    \caption{Direct estimates of the number of children, ages 0--1.}
    \label{fig:s48400direct}
  \end{subfigure}
  ~
  \begin{subfigure}{0.45\textwidth}
    \centering
    \includegraphics[width=\textwidth, height=2in]{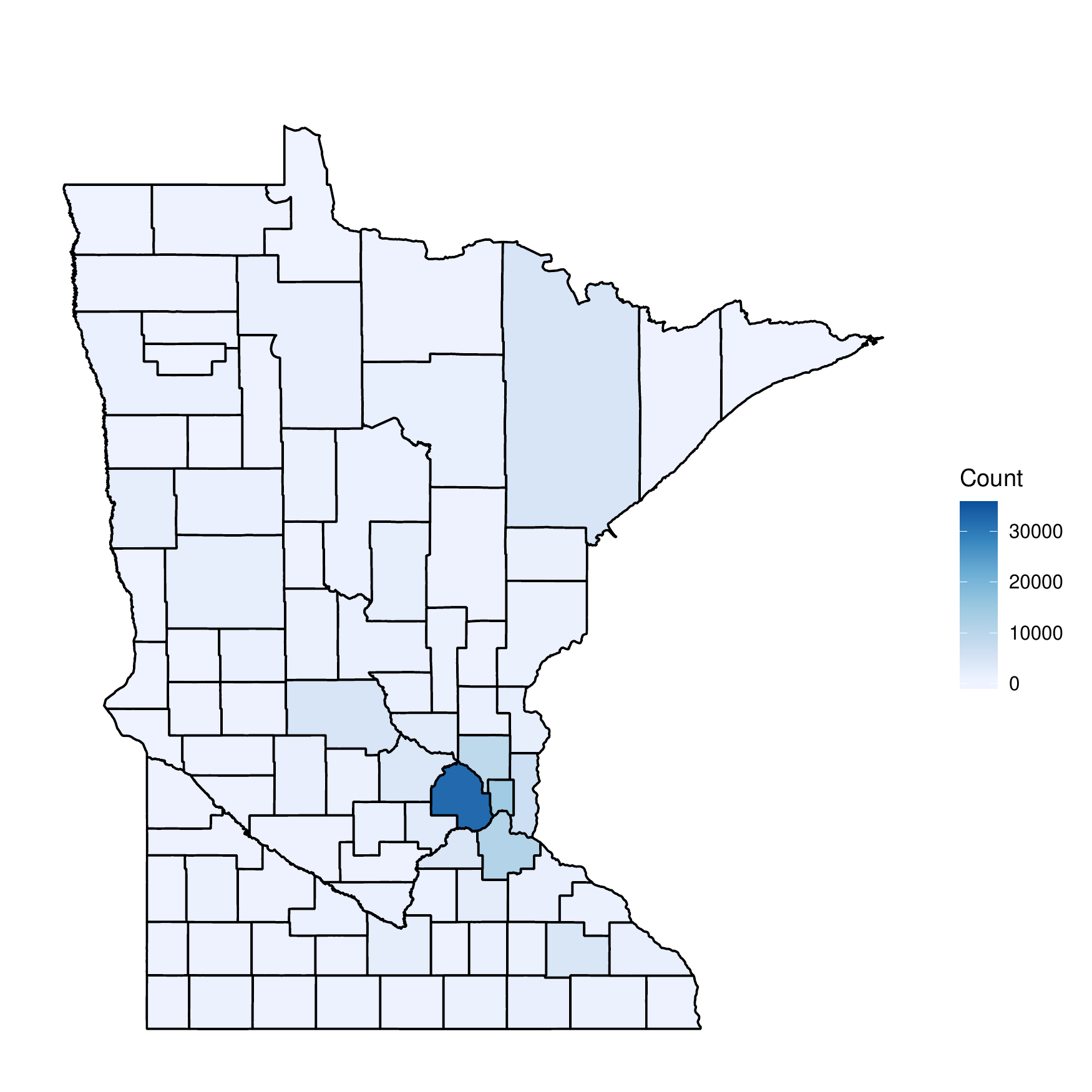}
    \caption{Model-based predictions of the number of children, ages 0--1.}
    \label{fig:s48400pred}
  \end{subfigure}

  \begin{subfigure}{0.45\textwidth}
    \centering
    \includegraphics[width=\textwidth, height=2in]{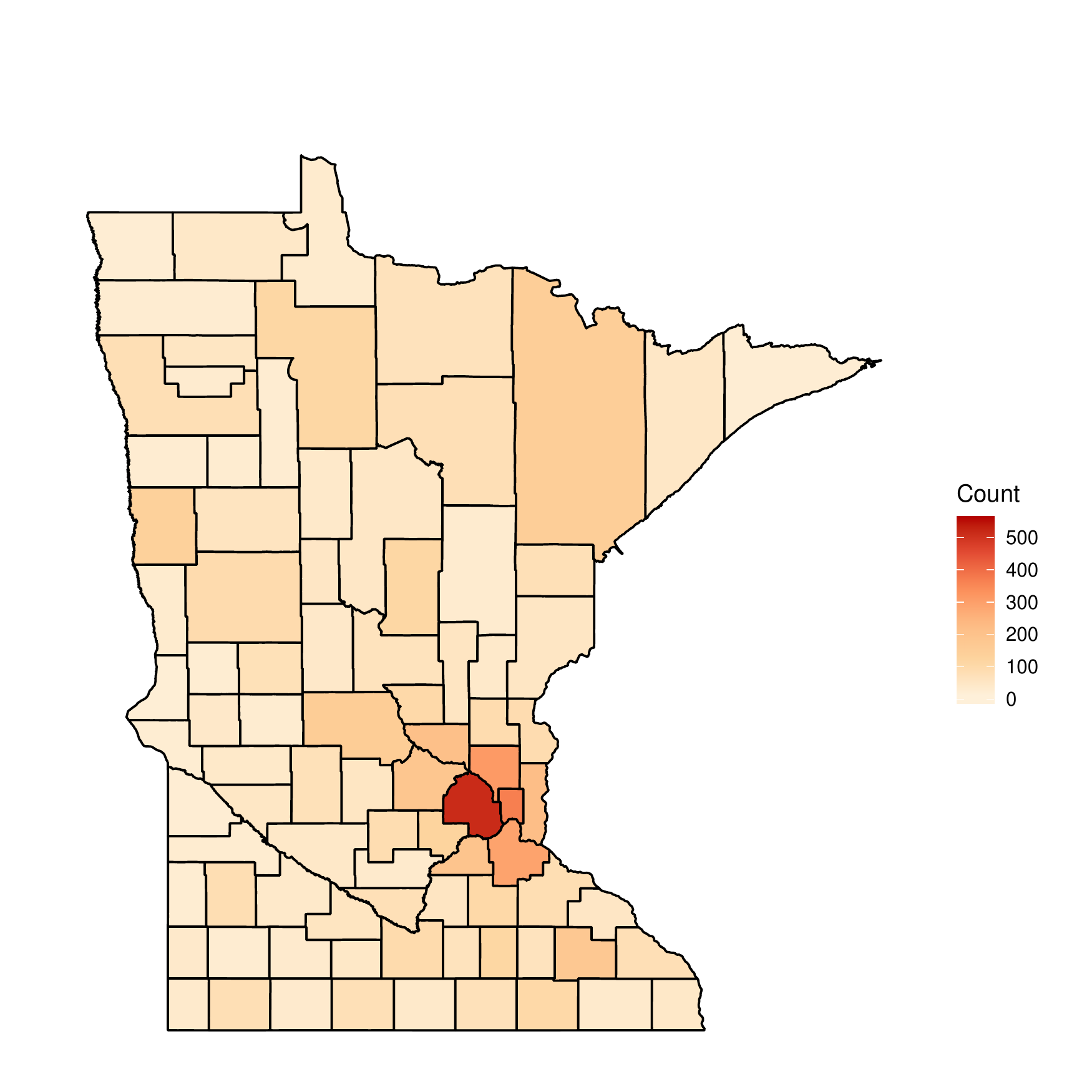}
    \caption{Standard errors of direct estimates of the number of children, ages 0--1.}
    \label{fig:s48400direct_se}
  \end{subfigure}
  ~
  \begin{subfigure}{0.45\textwidth}
    \centering
    \includegraphics[width=\textwidth, height=2in]{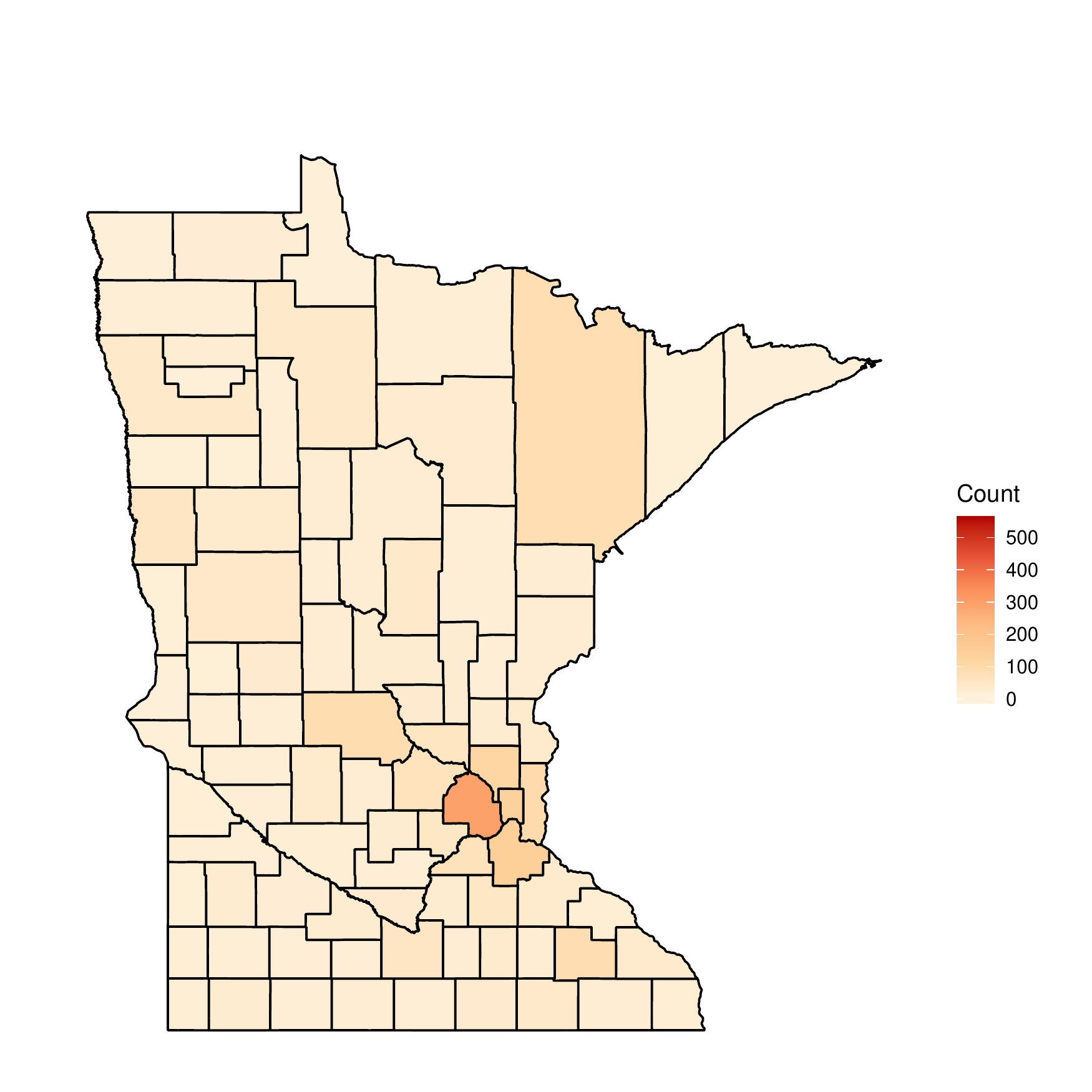}
    \caption{Posterior standard error of model-based predictions of the number of
      children, ages 0--1.}
    \label{fig:s48400pred_se}
  \end{subfigure}
  \caption{Comparison of the spatial patterns of the direct estimates and the
    predicted values, in blue, and a comparison of the spatial patterns of the
    standard errors of the direct estimates and predicted values, in
    red.}
  \label{fig:s48400all}
\end{figure}

The choice of the number of basis functions to use in the model specification remains an open
question.  Using too few basis functions can result in the model oversmoothing the data, while
increasing the number of basis functions can add additional variance and computational burden.
However, a careful choice of a reduced rank set of basis functions can be shown to produce
better predictions than those using a full set of basis functions \citep{BradleyEtAl2015}.  We
present results from fitting the model using 36 basis functions, which is approximately 50\% of
the available MI basis functions.  In a sensitivity study, we investigated the effects of
varying the number of basis functions used.  The predictions are relatively robust to changes
in \(r\), particularly when 50\% or more of the available MI basis functions are used.  When
too few basis functions are included, the model can over smooth the data, and predictions for
some of the small areas can have a higher bias.  The tradeoff with using more basis functions
is the increased computational burden of fitting the model.  With the datasets analyzed in this
project, a default of using 50\% of the available MI basis functions seems to be a sensible
choice.

\begin{figure}[t]
  \centering
  \includegraphics[height=3in, width=0.65\textwidth]{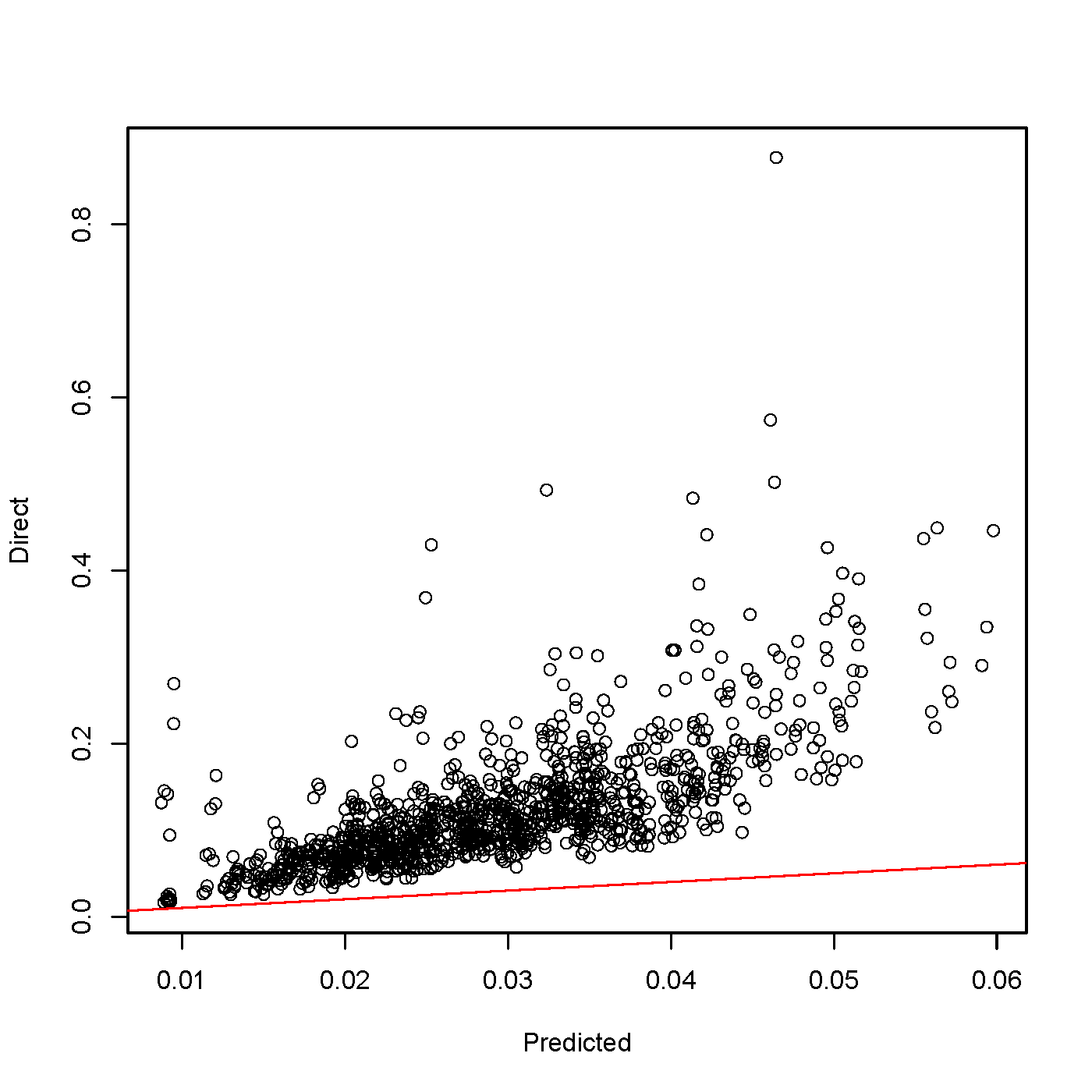}
  \caption{Comparison of the coefficients of variation of the predicted values
    vs. the coefficients of variation of the direct estimates, for all three margins:
    children ages 0--1, 2--3, and 4--5, for all counties in the Midwest. The diagonal
    line is shown in red.}
  \label{fig:s48400cv}
\end{figure}

We used the Stan modeling platform along with the RStan package \citep{sta18} in R
\citep{Rcore2019} to fit the multivariate spatial model to this ACS dataset.  Stan worked very
well in this example, as we were able to fit the model to this dataset using 2000 Hamiltonian
Markov chain (HMC) iterations, using the first 1000 iterations as burn-in,  in only a few
minutes using a Windows laptop (Intel i7-6600U CPU @ 2.60GZ, 16GB of RAM).  We used the
package's built-in diagnostics to monitor convergence, with no issues detected.

Figure~\ref{fig:s48400all} compares the direct estimates with the model-based predictions of
the number of children ages 0--1 in counties in Minnesota, and their associated standard
errors.  The direct estimates and model-based predictions are shown in blue in
Figures~\ref{fig:s48400direct} and \ref{fig:s48400pred}, respectively.  Here, it can be seen
that the spatial patterns of the point estimates are very similar.  Importantly, the
model-based predictions are `close' to the direct estimates in areas with large sample sizes,
so that the model-based predictions preserve the direct estimates with low sampling variance.
The predictions mainly differ from the direct estimates in areas with smaller sample size.  In
the areas with smaller sample size, the multivariate spatial model utilizes the spatial and
multivariate correlation, to `borrow information' across and within areas.  A comparison of the
standard errors of the direct estimates and model-based predictions is presented in
Figures~\ref{fig:s48400direct_se} and \ref{fig:s48400pred_se}.  Here, as was seen with the
point estimates, the spatial pattern of the standard errors is maintained.  We also see
significantly reduced standard errors in Figure~\ref{fig:s48400pred_se}, compared to
Figure~\ref{fig:s48400direct_se}.

Figure~\ref{fig:s48400cv} compares the coefficients of variation of the direct estimates with
the coefficients of variation of the model-based predictions for all three margins: children
ages 0--1, 2--3, and 4--5, for all counties in the Midwest.  Here, we see a drastic increase in
the precision of the model-based predictions over the corresponding direct estimates, with an
overall average reduction in the coefficients of variation of approximately 73\%.  The
improvement in precision, in this example, is uniform, with all counties seeing a reduction in
the coefficient of variation of the model-based predictions.  However, the most dramatic
improvements in precision tend to be in the counties with smallest sample size.  This is the
`borrowing of strength' phenomenon, which is often seen in small area estimation problems,
where the effective sample size in small areas is increased by utilizing information from
larger areas, thereby increasing precision of estimates.  The preservation of spatial patterns
of model-based predictions, along with an increase of precision of these point estimates has
important policy consequences, as there is potential for more, and higher quality data
releases, at possibly finer levels of geography than are currently available.

\subsection{Estimation of the number of children in counties in Minnesota by
  race}\label{sec:msmfail}

\begin{figure}[t]
  \centering
  \includegraphics[height=3in,width=0.65\textwidth]{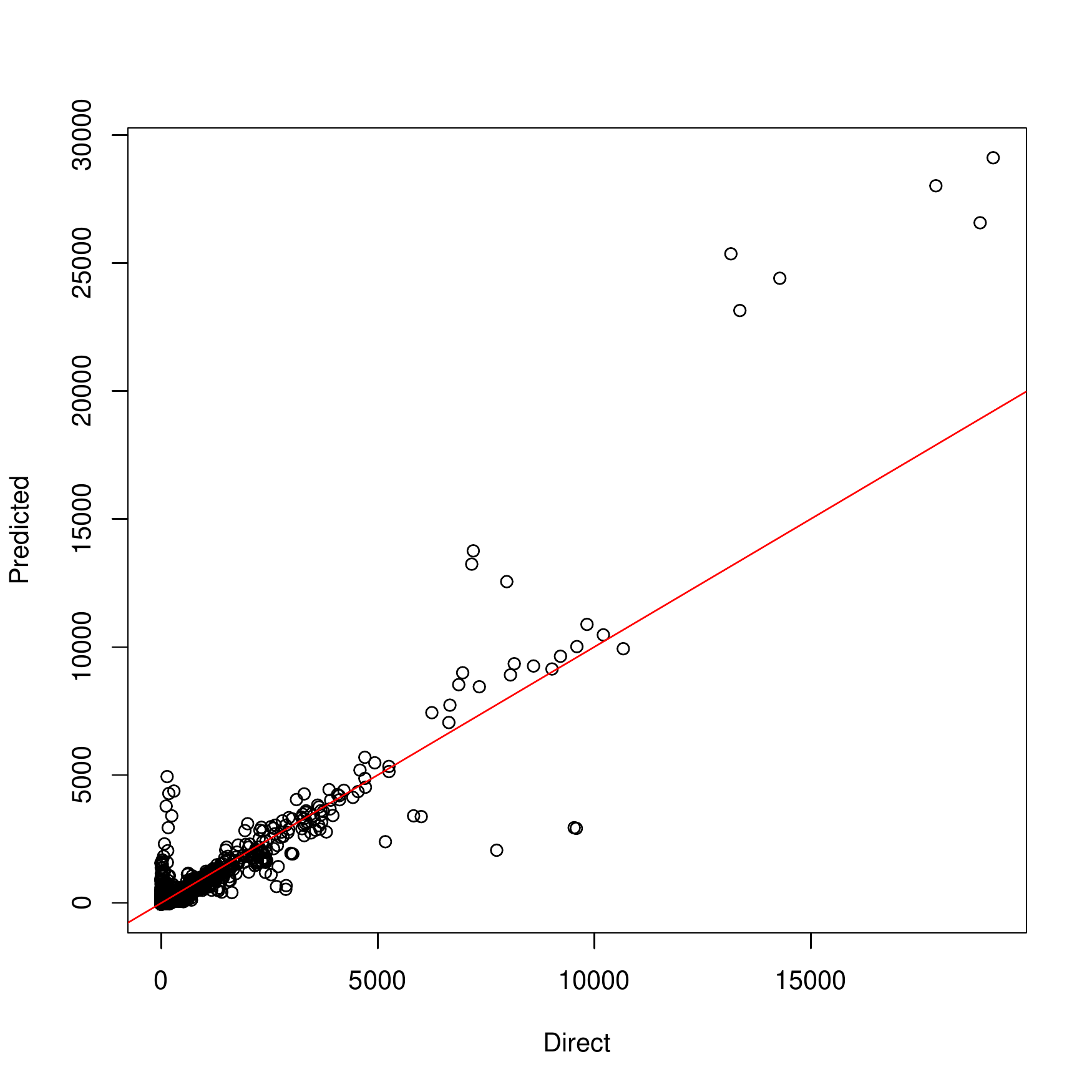}
  \caption{Comparison of the direct estimates and model-based estimates of the number of
    children in the 21 age by race groups in counties in Minnesota.}
  \label{fig:msmFail}
\end{figure}

We now fit the multivariate spatial model to direct estimates of the number of children, ages
0--1, 2--3, and 4--5 cross-classified by seven race categories, White Alone, Black Alone, Asian
Alone, American Indian or Alaska Native Alone, Native Hawaiian or Pacific Islander alone, Other
alone, or two or more races, in counties in Minnesota.  The model specification and
computational details, including covariates, priors, and MCMC methods, was the same as those
used in the analysis in Section~\ref{sec:s48400}.  Also, as in Section \ref{sec:s48400}, we fit
the model to direct estimates associated with counties in Minnesota, as well as surrounding
Midwestern states.

Figure \ref{fig:msmFail} shows a scatter plot of the predicted values in each of the 21
different age by race groups for each of the counties in Minnesota, against the corresponding
direct estimates.  Clearly the predicted values in Figure \ref{fig:msmFail} are unacceptable,
as some of the predicted values deviate from the direct estimates wildly.  At the upper right
of Figure \ref{fig:msmFail}, we see predicted values of White children in some counties
are nearly double the direct estimates.  Because these direct estimates use the large sample
sizes for counties containing large cities and are expected to be relatively precise, we would
expect the associated predicted values to closely match the direct estimates.  On the other
end, in the lower left of Figure \ref{fig:msmFail}, we see direct estimates that are near zero,
but with corresponding predicted values as much as 5,000, which in some cases is
larger than the total number of people in that county.  These observations taken together
suggest large biases in the predicted values.

\begin{figure}[t]
  \centering
  \begin{subfigure}{0.45\textwidth}
    \centering
    \includegraphics[width=\textwidth,height=2in]{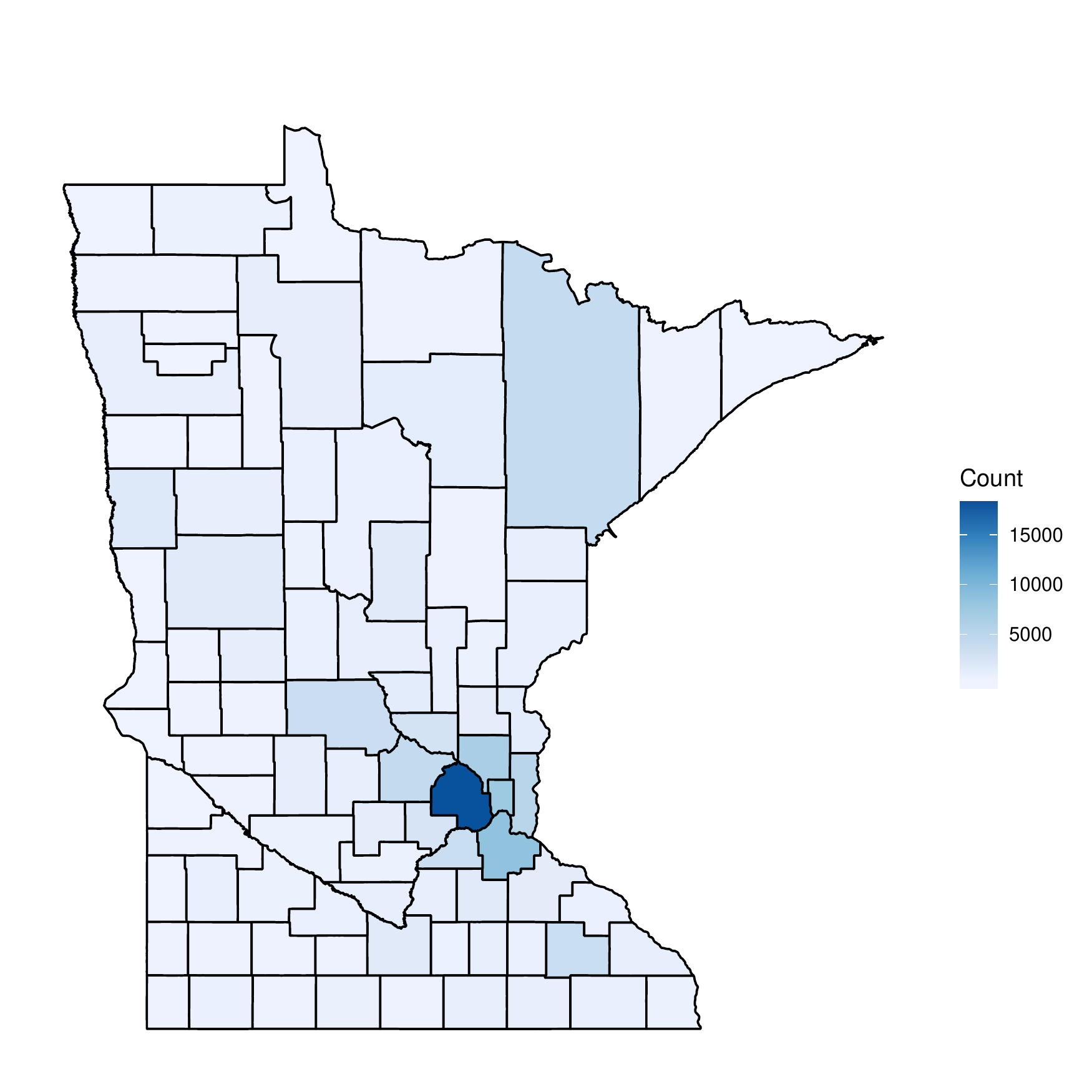}
    \caption{Direct estimates of the number of White children, ages 2--3.}
    \label{fig:s48405direct1}
  \end{subfigure}
  ~
  \begin{subfigure}{0.45\textwidth}
    \centering
    \includegraphics[width=\textwidth,height=2in]{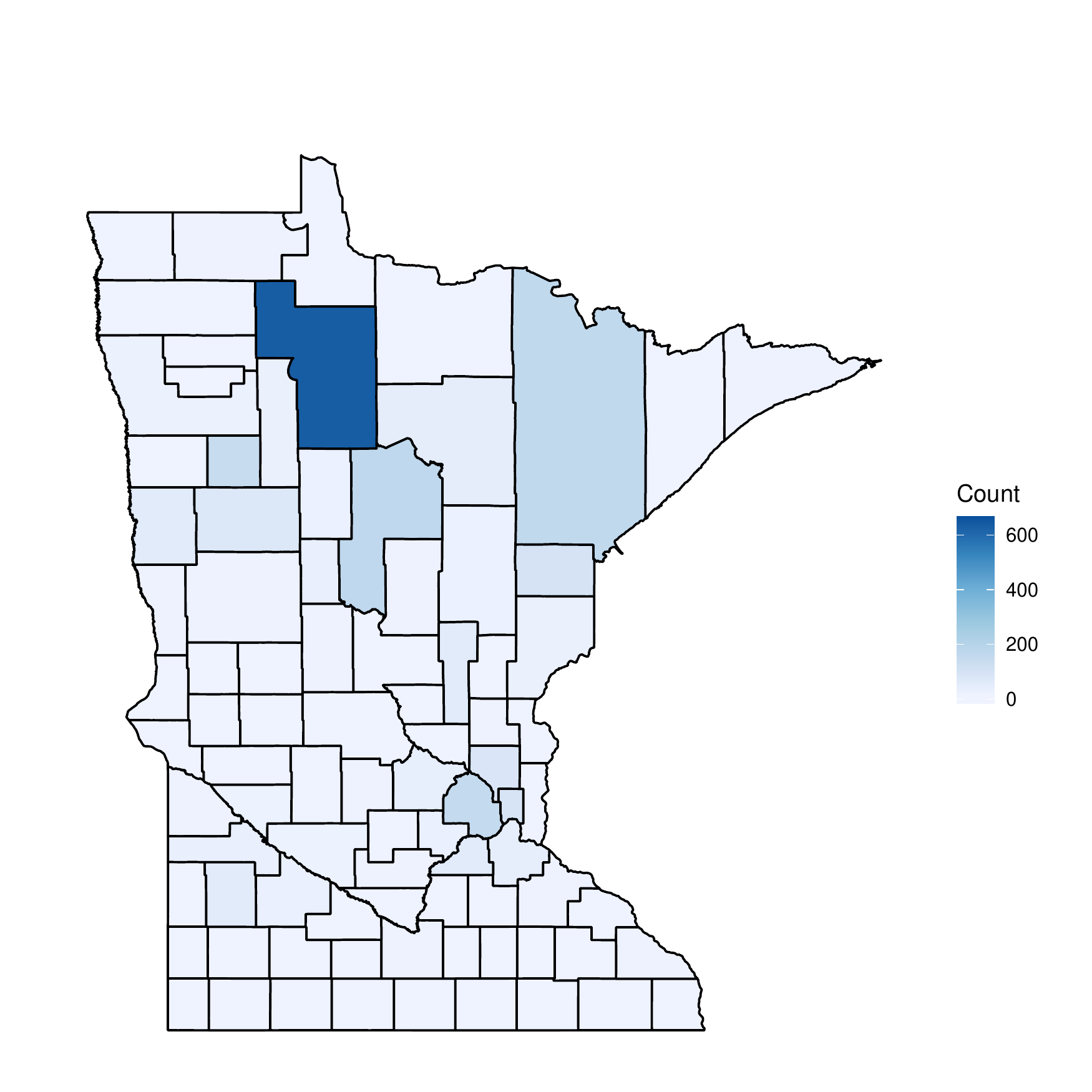}
    \caption{Direct estimates of the number of American Indian or Alaska Native
      children, ages 2--3.}
    \label{fig:s48405direct2}
  \end{subfigure}
  \caption{A comparison of the direct estimates of the number of White children, ages 2--3,
    with the direct estimates of the number of American Indian or Alaska Native children, ages
    2--3, showing the varying spatial patterns on different margins within the multivariate
    data.}
  \label{fig:s48405both}
\end{figure}

The reason for the large apparent biases in the predicted values can be inferred by looking at
spatial plots of different margins of the tabulated ACS data.  Figure \ref{fig:s48405both}
shows spatial plots of the age by race direct estimates for counties in Minnesota for two
margins.  Figure \ref{fig:s48405direct1} shows the direct estimates of the number of White
children, ages 2--3, and Figure \ref{fig:s48405direct2} shows the direct estimates of the
number of American Indian or Alaska Native children, ages 2--3.  Clearly the spatial patterns
in these two figures are quite different, with the largest number of White children estimated
in to be in Hennepin County, which includes Minneapolis, while the greatest number of American
Indian or Native American children are estimated to be in the more rural, Beltrami County, in
Northern Minnesota.  In addition to the different spatial patterns, the scale of the margins of
the data are very different, with the direct estimates of the number of White children ranging
from 0--150,000, while the direct estimates of the number of American Indian or Alaska Native
children range from 0--600, in this dataset.

The varying spatial patterns shown in Figure \ref{fig:s48405both}, and the resulting model
assumption violation, help explain the poor performance of the predicted values from the fitted
multivariate spatial model in this example.  The predicted values aggressively smooth the
direct estimates based on a common assumed spatial field.  The shared spatial random effects
for the different margins of the data result in large apparent biases, as the fitted model
seems to try to compromise between the different spatial patterns, multivariate characteristics
of the data, and the different scales of the direct estimates.  Reexamination of Figure
\ref{fig:msmFail} does suggest a clustering effect, with predictions falling far from the
diagonal, appearing to cluster in 3 or 4 groups.

Exploiting the spatial dependence, as well as the multivariate dependence, present in the data
by incorporating a multivariate spatial process into the model is critical for producing
predictions which have reduced standard errors from the corresponding direct estimates.  In
separate analyses (not shown) in which independence models were fit to ACS direct estimates, we
did not observe meaningful increases in precision.  However, a poorly specified process model
with a common assumed spatial field, as appears to be the case in this example, results in
predictions with potentially serious biases for certain areas.  One strategy might be to expand
the number of basis functions to allow more flexibility in the model to accommodate the
different aspects of the data.  However, in this example, this approach was unsuccessful.  A
second strategy is to subset the data into more homogeneous groups, and to fit the multivariate
spatial model to each subset.  The problem here is that it is not always apparent how to subset
the data, particularly as the dimension of the data and the number of cross-classifications
present in the special tabulations increase.  It is more desirable to have a data-driven
method to cluster the data for these high dimensional problems.  In the next section we extend
the multivariate spatial model, by introducing a mixture component into the process model, to
accommodate more complicated datasets with varying multivariate spatial characteristics.

\section{Multivariate Spatial Mixed Effects Model with Dirichlet Process Mixing}
\label{sec:msmmix}

For analysis of multivariate datasets with potentially varying spatial patterns, it is of
interest to develop model-based methods which can cluster the observed data according to common
multivariate characteristics and common spatial patterns, in addition to producing precise
area-level predictions.  Also, for the reasons previously discussed, we may not have strong
prior information about the number of clusters.  It is therefore desirable to allow for
uncertainty in the number of clusters that are used.  The Dirichlet process \citep{fer73}
naturally incorporates these two properties.

For more complicated datasets, we introduce the following extension of the multivariate mixed
effect spatial model, which incorporates Dirichlet process mixing on the latent Gaussian
process and regression coefficients.  As before, the data model is given by
\begin{equation*}
  Z^{(l)}(A) = Y^{(l)}(A) + \varepsilon^{(l)}(A),
\end{equation*}
for \(l = 1, \dots, L\) and \(A \in \mathcal{D}\). Writing $\boldsymbol{\theta}^{(l)}(A)^\top =
\left( \boldsymbol{\beta}^{(l)}(A)^\top, \boldsymbol{\eta}^{(l)}(A)^\top \right)$, the process
model is now
\begin{equation}\label{E:msmix}
\begin{gathered}
  Y^{(l)}(A) = \boldsymbol{x}^{(l)}(A)^\top \boldsymbol{\beta}^{(l)}(A) +
    \boldsymbol{\psi}^{(l)}(A)^\top \boldsymbol{\eta}^{(l)}(A) \\
  \boldsymbol{\theta}^{(l)}(A) \mid G \sim G \\
  G \mid \alpha, G_0 \sim \text{DP} ( \alpha G_0).
\end{gathered}
\end{equation}
The Dirichlet process (DP) prior is used as the clustering mechanism in the process model.  The
parameter \(\alpha\) is an unknown concentration parameter controlling the degree of
clustering. The measure \(G_0\) is a `base' measure on \(\boldsymbol{\theta}^{(l)}(A)\), which
is assumed known, up to a finite-dimensional parameter.  We specify \(G_0\) to be a product of
Gaussian distributions, \(\text{N}_p (\boldsymbol{0}, \sigma^2_{\boldsymbol{\beta}} \boldsymbol{I}_{p \times
  p})\) and \(\text{N}_r (\boldsymbol{0}, \sigma^2_{\boldsymbol{\eta}} \boldsymbol{K})\), on the
\(\boldsymbol{\beta}\) and \(\boldsymbol{\eta}\) components of \(\boldsymbol{\theta}\),
respectively.

To complete the model specification, we need to choose prior distributions on the unknown
parameters \(\sigma^2_{\boldsymbol{\eta}}\) and \(\alpha\).  We let
\(\sigma^2_{\boldsymbol{\eta}} \sim \text{IG} ( a_{\boldsymbol{\eta}},
b_{\boldsymbol{\eta}})\), and \(\alpha \sim \text{Gamma} ( a_\alpha, b_\alpha )\), for fixed
hyperparameters \(a_{\boldsymbol{\eta}}, b_{\boldsymbol{\eta}}, a_\alpha\), and \(b_\alpha\).
We generally set \(a_{\boldsymbol{\eta}}\), \(b_{\boldsymbol{\eta}}\), \(a_\alpha\) and
\(b_\alpha\) to be small, positive constants, and \(\sigma^2_{\boldsymbol{\beta}}\) to be a
large number, giving vague, but proper priors on \(\boldsymbol{\beta}\), \(\alpha\), and
\(\sigma^2_{\boldsymbol{\eta}}\).

This choice of parameter model, when combined with the data augmentation approach of
\citet{esc95}, gives full conditional distributions that are all from simple parametric
families.  Gibbs sampling can then be used to sample from the posterior distribution, using
methods from \citet{nea00}.  Computational details can be found in the Supplementary Materials.

\begin{remark}
  While Gibbs sampling algorithms exist for the nonparametric Dirichlet process
  prior, the computational burden can be large for two main reasons.  First, because the
  number of clusters is random in this model specification, there is potentially a large memory
  need, as parameter values need to be drawn and saved for each cluster, at each iteration of
  the Gibbs sampler.  In particular, at early stages of the Gibbs sampler, there are often a
  large number of clusters that are investigated prior to convergence of the MCMC chain.
  Second, the number of iterations for convergence of the Gibbs sampler can be large,
  especially if the initial values of the unknown parameters are chosen poorly.

  To speed up convergence, a parametric approximation to the Dirichlet process prior can be
  used.  The Dirichlet process with base measure \(G_0\) and concentration parameter \(\alpha\)
  can alternatively be written as the random measure
  \begin{equation}\label{E:stick}
    \mathcal{P} ( \cdot ) = \sum^\infty_{k = 1} \pi_k \delta_{Z_k} ( \cdot ),
  \end{equation}
  known as the stick-breaking representation \citep{set94}. Here, the \(\delta_{Z_k} ( \cdot
  )\) are point masses concentrated on random variables \(Z_k \iid G_0\), and the \(\pi_k = V_k
  \prod_{b = 1}^{k-1} (1 - V_b) \) are weights based on random variables \(V_k \iid
  \text{Beta}(1, \alpha)\).

  If there is a priori knowledge of a maximum number of clusters, the series \eqref{E:stick}
  can be truncated to a number of terms, say \(M\), which is greater than the known number of
  clusters.  The truncated random measure
  \begin{equation}\label{E:trunc}
    \mathcal{P}_M ( \cdot ) = \sum^M_{k = 1} \pi_k \delta_{Z_k} ( \cdot ),
  \end{equation}
  can be used as a parametric approximation to the Dirichlet process prior, and inference can
  be done using a Gibbs sampler, described in detail in \citet{ish01}.  \xqed{$\blacksquare$}
\end{remark}

\begin{remark}
  In the experiments done in this paper, inference using the parametric approximation in
  \eqref{E:trunc} was very similar to inference using the exact Dirichlet process prior in
  \eqref{E:stick}, so long as the truncation level \(M\) was sufficiently large.  If the
  maximum number of clusters, \(M\), was chosen too small, we found the multivariate spatial
  mixture model, using the truncated measure \eqref{E:stick} as a prior, had some of the same
  problems as the multivariate spatial model as was seen in Section~\ref{sec:msmfail}.  With
  the datasets studied in this paper, setting \(M = 25\) worked well.  The major advantage of
  using the prior \eqref{E:trunc} over the Dirichlet process prior is that the computational
  burden is greatly reduced.  Both the memory requirements, and the computing time for
  convergence of the Gibbs sampler was much less, making use of \eqref{E:trunc} potentially
  more appealing for production of official estimates, particularly as the size of datasets and
  the number of datasets to analyze can be quite large. \xqed{$\blacksquare$}
\end{remark}

\begin{remark}
  While our main focus is on producing area-level predictions with higher precision than the
  corresponding direct survey estimates, we are also interested in investigating the
  effectiveness of the clustering mechanism of the nonparametric spatial mixture model, and the
  degree to which the added model uncertainty of an unknown number of clusters affects the
  precision of the model-based predictions.  The numerical example and data analysis in the
  following sections present results which use the multivariate spatial mixture with the
  nonparametric Dirichlet process \eqref{E:stick} as a prior. \xqed{$\blacksquare$}
\end{remark}

\section{Empirical Simulation Study}
\label{sec:empirical}

In Section~\ref{sec:msmfail}, we showed that the multivariate spatial model can produce
predictions of areal quantities that are of obvious poor quality when there are varying spatial
and multivariate patterns.  We proposed an extension of the multivariate spatial model in
Section~\ref{sec:msmmix}, which clusters the data on these varying characteristics, by introducing
a Dirichlet process prior on the process model.  In this section we present results of a
data-based, empirical simulation study, designed to study properties of this multivariate
spatial mixture model (MSMM).  The two main properties of interest are, first, to verify that
the MSMM can effectively cluster data based on spatial patterns and multivariate properties of
the data, and second, that the predictions at the areal level are more precise, on average,
than the corresponding direct estimates.  For comparison, we also show the performance of
predictions using a Fay-Herriot model \citep{fay79}, given by
\begin{equation}\label{E:fay}
\begin{gathered}
  Z^{(l)} (A) = Y^{(l)} (A) + \varepsilon^{(l)} (A) \\
  Y^{(l)} (A) = \boldsymbol{x}^{(l)} (A)^\top \boldsymbol{\beta} + \nu^{(l)}(A), 
\end{gathered}
\end{equation}
where \(\varepsilon^{(l)}(A) \indep \text{N}(0, D^{(l)}(A))\) and \(\nu^{(l)}(A) \iid
\text{N}(0, \sigma^2)\).  The Fay-Herriot model \eqref{E:fay} does not take into consideration
any multivariate dependence or spatial dependence in the data, except for any information
included in the covariates \(\boldsymbol{x}^{(l)}(A)\).

\begin{remark}
  The multivariate spatial model, introduced in Section \ref{sec:MSM}, was also fit to the
  simulated datasets.  However, the performance of the predictions using this model are not
  shown, as these predictions suffer from the same problems as those discussed in
  Section~\ref{sec:msmfail}.  Primarily, the bias of the predictions in some areas is huge,
  making this model inappropriate for use with this type of data. \xqed{$\blacksquare$}
\end{remark}

The simulation study is designed around the age by race dataset, discussed in Section
\ref{sec:msmfail}, which could not be effectively modeled by the multivariate spatial model.
Let \(Z^{(l)} (A)\) represent the log of the direct ACS 5-year estimates of the counts, plus 1,
given in equation~\eqref{E:logDirect}, for each age by race cross-classification \(l = 1,
\dots, 21\), in each of the counties, \(A \in \mathcal{D}\), in Minnesota and surrounding
states.  Let \(D^{(l)} (A)\) be the sampling variance of the \(Z^{(l)} (A)\).

\begin{remark}
  The direct estimates of the counts, as in Equation~\eqref{E:direct}, and their associated
  direct variance estimates are publicly available data.  The U.S. Census Bureau uses the
  successive differences replication method \citep{jud90, fay95, tor14} to create replicate
  weights for variance estimation.  These replicate weights can also be used to estimate the
  variance of the direct estimates of the log counts; these are the variance estimates that
  are used in Section \ref{sec:s48405}.  However, the variance estimates of the log counts are
  not publicly available data.  In order to present results of a numerical example based
  completely on publicly available data, we instead use the delta method to transform the
  variance estimates from the count scale to the log count scale, and then smooth the
  transformed estimates so they more closely agree with the replicate weight variance
  estimates.  Details can be found in the Supplementary Materials. \xqed{$\blacksquare$}
\end{remark}

The perturbed version of the log counts is
\begin{equation}\label{E:simsetup}
  R^{(l)} (A) = Z^{(l)} (A) + \varepsilon^{(l)} (A), \ l = 1, \dots, 21, \ A \in
    \mathcal{D},
\end{equation}
where \(\varepsilon^{(l)} (A) \indep \text{N}(0, D^{(l)} (A))\).  In this setup, the \(D^{(l)}
(A)\) are used as the sampling variance of the log counts, \(R^{(l)}(A)\), and are assumed
known.  For the purpose of this empirical study, we act as if the direct estimates of the log
counts, \(Z^{(l)} (A)\), are the unobserved, true multivariate spatial latent process, and
treat the \(R^{(l)} (A)\) as the data process.  This empirical simulation study is similar to
what is done in \citet{BradleyEtAl2015} and \citet{BradleyEtAl2018}, and is designed as a way
of generating data that behave similar to what might be observed in practice.

We generate 100 datasets from \eqref{E:simsetup}, giving us ``observed'' values \(\left\{
\left(R^{(l)}(A), D^{(l)}(A) \right) \right\}\).  The MSMM is fit to the perturbed values
\(R^{(l)} (A)\) in each simulated dataset to predict the \(Z^{(l)} (A)\).
The covariates used in the model include an intercept, the log of the total county
population (which is assumed known from Census data), and a collection of dummy variables
corresponding to the different age by race cross-classifications.  The hyperparameters used
were \(a_{\boldsymbol{\eta}} = b_{\boldsymbol{\eta}} = 0.1\), \(\sigma^2_{\boldsymbol{\beta}} =
100\), \(a_\alpha = 1\) and \(b_\alpha = 4\).  These hyperparameters give vague, but proper
priors, on \(\boldsymbol{\beta}\) and \(\sigma^2_\eta\).  Since, with the Dirichlet process,
the number of clusters is asymptotically equal to \(\alpha \log n\) \citep{kor73}, we chose the
hyperparameters \(a_\alpha = 1\) and \(b_\alpha = 4\) put prior mass on smaller values of
\(\alpha\), with 90\% of the prior mass on $(0.2, 1)$, giving prior preference to a smaller
number of clusters.  We investigated different values of \(a_\alpha\) and \(b_\alpha\), and did
not find much sensitivity to the choice of hyperparameters, except for the amount of time to
reach convergence, and the number of clusters which were created at early iterations of the
Gibbs sampler.

For each simulated dataset $i = 1, \ldots, 100$, the MSMM was fit to the perturbed data,
\(\{R^{(l)}_i(A)\}\) using a Gibbs sampler whose derivation and computational details can be
found in the Supplementary Materials.  The sampler was run for 5,000 iterations, with a burn-in
of 1,000 iterations.  Assessing convergence of the Gibbs sampler is challenging in the mixture
model framework, due to potential label-switching.  At each iteration of the MCMC
chain, the number of clusters can change, and the associated cluster labels are arbitrary and
change from iteration to iteration.  We can, however, monitor the MCMC chains of parameters
that are not dependent on the cluster labels, such as the concentration parameter, \(\alpha\),
the variance parameters, \(\sigma^2_{\boldsymbol{\beta}}\) and
\(\sigma^2_{\boldsymbol{\eta}}\), as well as predicted values of areal quantities which are
invariant to label-switching.  From this assessment, there was no lack of convergence detected,
based on visual inspection of the MCMC chains for these parameters.  Additionally, batch means
\citep{jon06}, using the square root rule, and the Geweke statistic \citep{gew92}, were
computed as formal tests of MCMC convergence. Neither of these diagnostics indicated lack of
convergence.

\begin{figure}[t]
  \centering
  \begin{subfigure}{0.32\textwidth}
    \centering
    \includegraphics[height=1.5in, width=\textwidth]{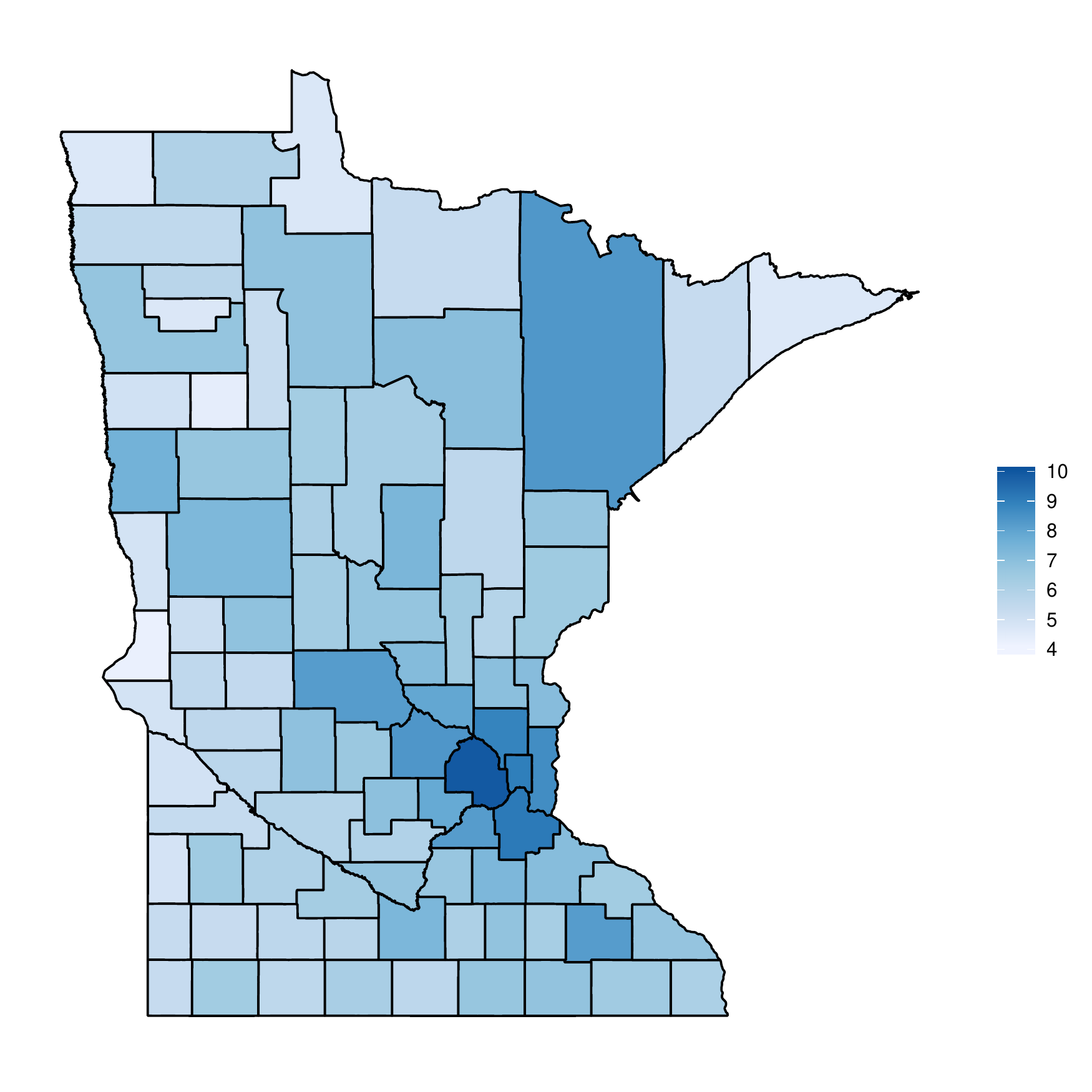}
    \caption{``True'' values of the log count of White children, ages 2--3}
    \label{fig:simNew_4_map_dir}
  \end{subfigure}
  ~
  \begin{subfigure}{0.32\textwidth}
    \centering
    \includegraphics[height=1.5in, width=\textwidth]{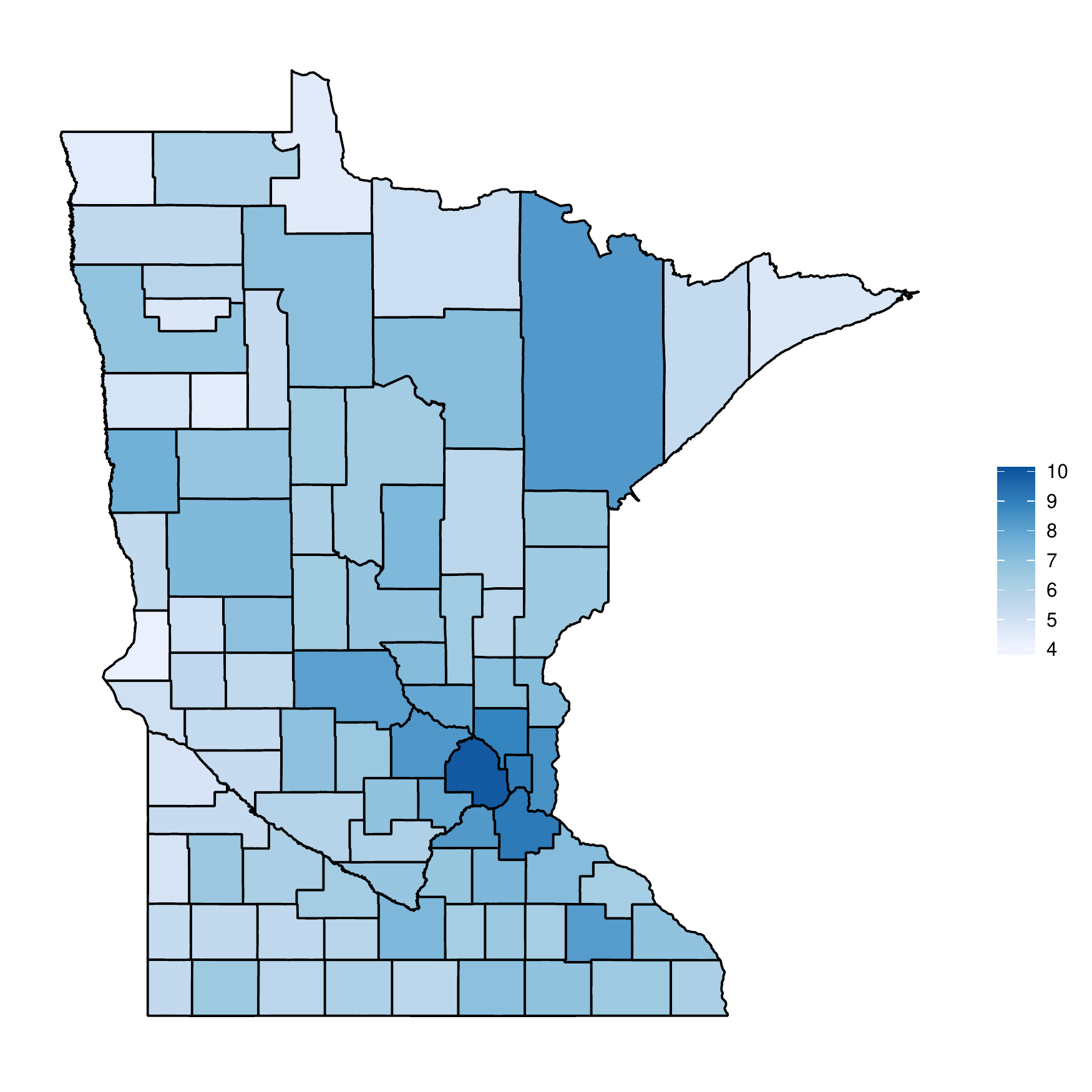}
    \caption{Direct estimates of the log count of White children, ages 2--3}
    \label{fig:simNew_4_map_pseudo}
  \end{subfigure}
  ~
  \begin{subfigure}{0.32\textwidth}
    \centering
    \includegraphics[height=1.5in, width=\textwidth]{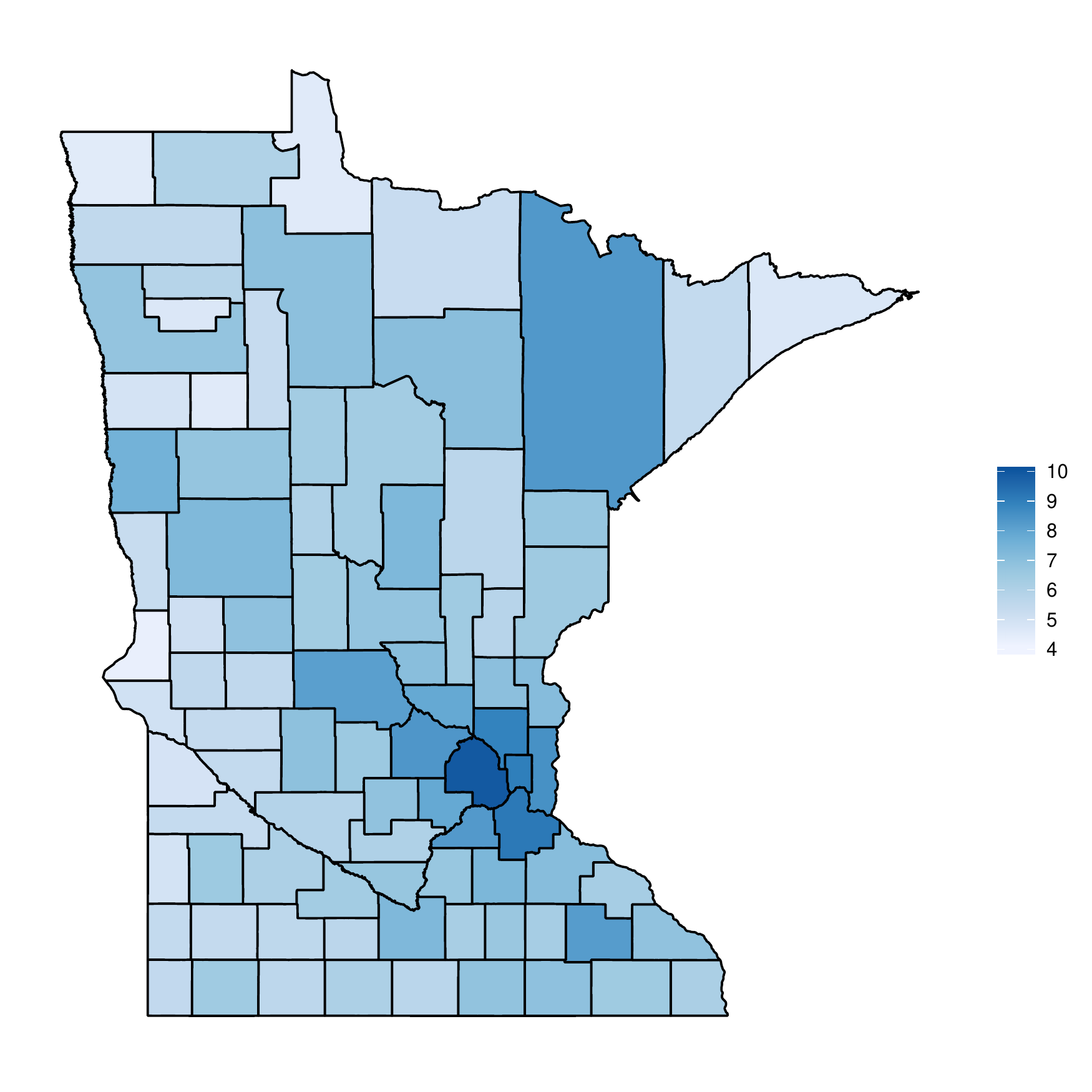}
    \caption{Model-based predictions of the log count of White children, ages 2--3}
    \label{fig:simNew_4_map_pred}
  \end{subfigure}

  \begin{subfigure}{0.32\textwidth}
    \centering
    \includegraphics[height=1.5in, width=\textwidth]{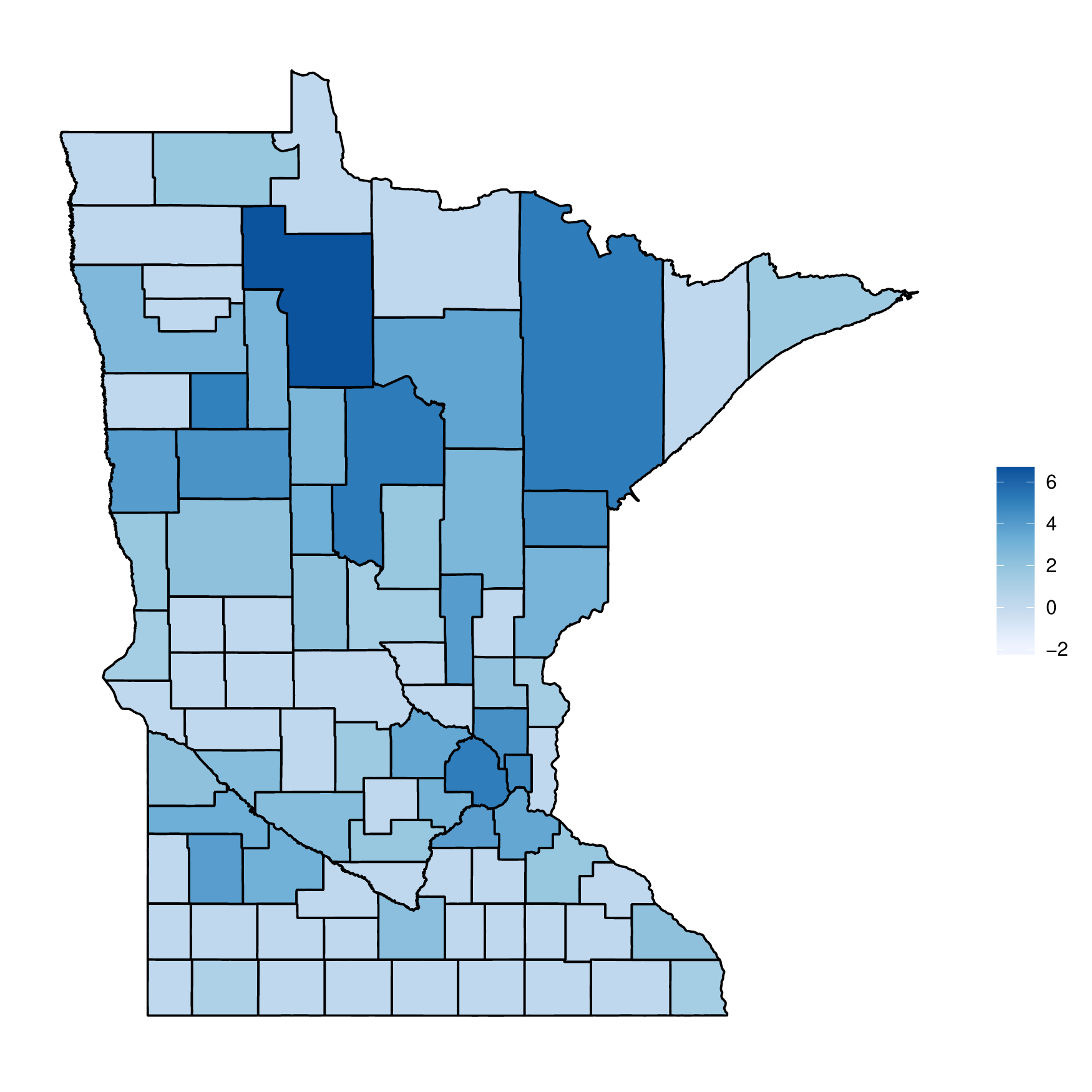}
    \caption{``True'' values of the log count of American Indian or Alaska Native
      children, ages 2--3.}
    \label{fig:simNew_12_map_dir}
  \end{subfigure}
  ~
  \begin{subfigure}{0.32\textwidth}
    \centering
    \includegraphics[height=1.5in, width=\textwidth]{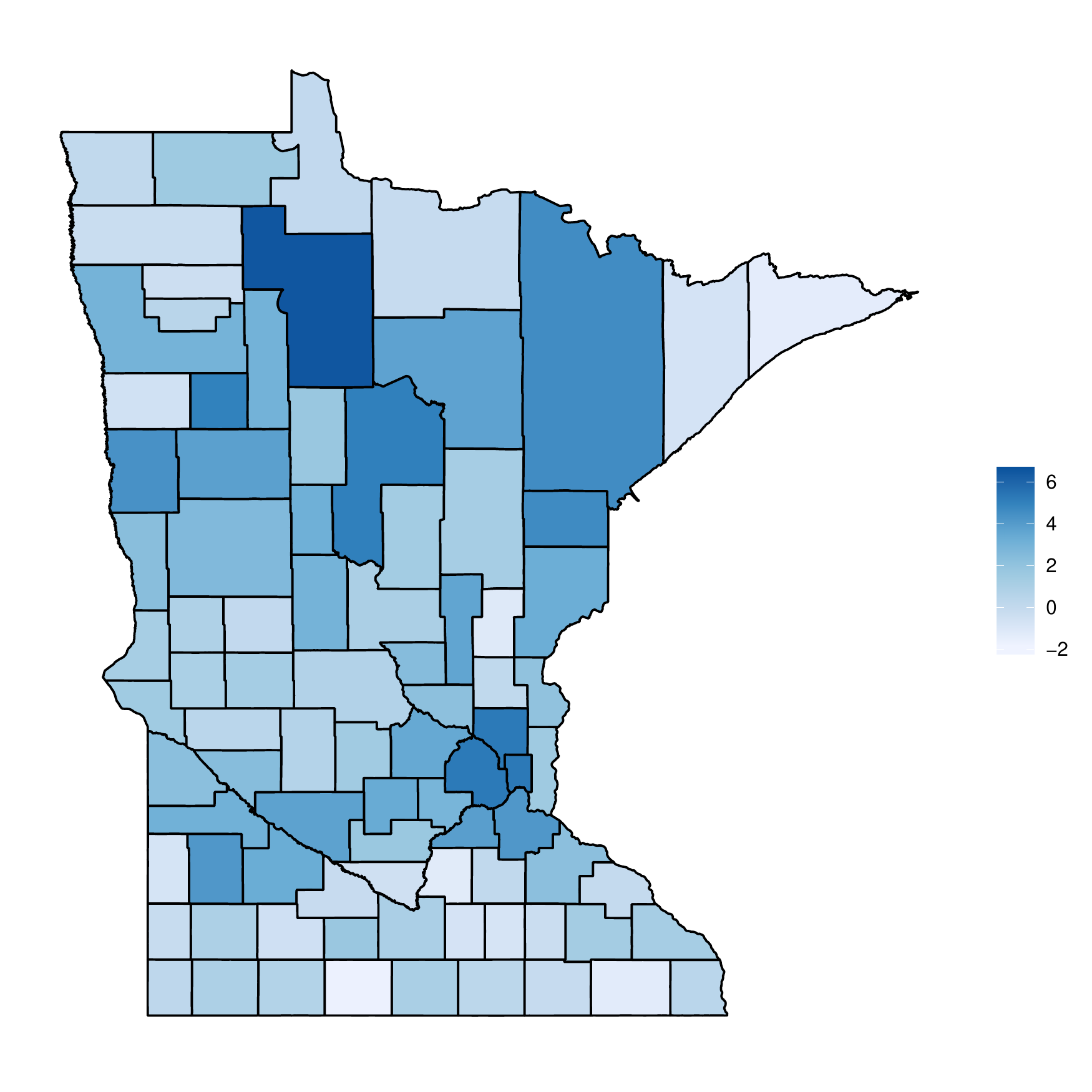}
    \caption{Direct estimates of the log count of American Indian or Alaska Native
      children, ages 2--3.}
    \label{fig:simNew_12_map_pseudo}
  \end{subfigure}
  ~
  \begin{subfigure}{0.32\textwidth}
    \centering
    \includegraphics[height=1.5in, width=\textwidth]{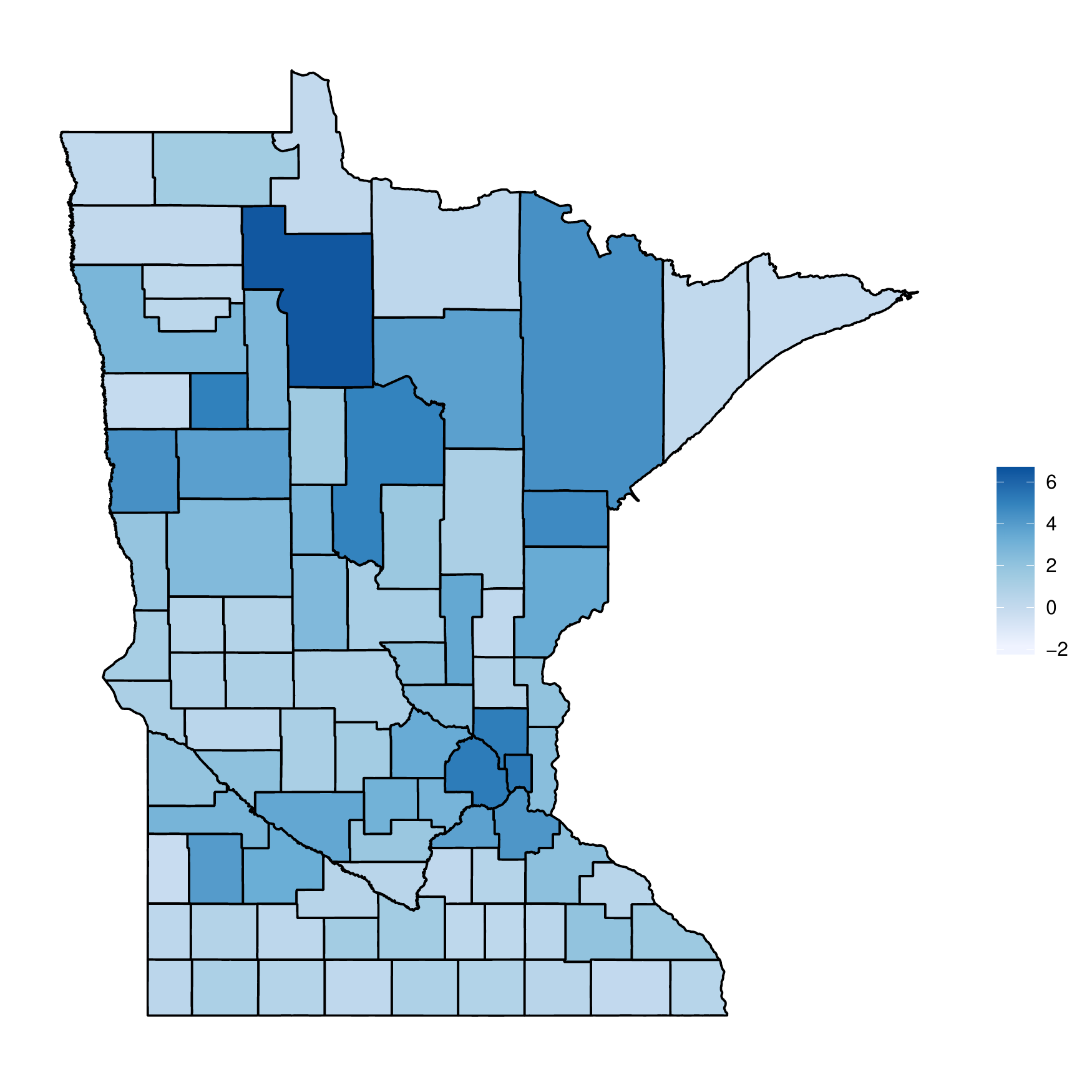}
    \caption{Model-based predictions of the log count of American Indian or Alaska
      Native children, ages 2--3.}
    \label{fig:simNew_12_map_pred}
  \end{subfigure}
  \caption{True values of the log count, perturbed values of the log count, and model based
    predictions of the log count, using the multivariate spatial mixture model.  The first row
    shows these quantities for White children, ages 2--3, and the second row shows these
    quantities for American Indian or Alaska Native children, ages 2--3, in counties in
    Minnesota.}
  \label{fig:sim_map_point}
\end{figure}

Let \(\{\hat{Z}^{(l)}_i (A)\}\) denote the model-based predicted values of the log counts,
\(Z^{(l)}(A)\), from fitting the MSMM model to the \(i\)th simulated dataset for \(i = 1,
\dots, 100\).  In Figure \ref{fig:sim_map_point} we show the ``true'' values of the log counts
for counties in Minnesota for two margins: the log of the number of White children, ages 2--3,
and the log of the number of American Indian or Alaska Native children, ages 2--3.  Also shown
are the perturbed data from the first simulation, as well as the predicted values of the log
counts using the MSM mixture model fit to this perturbed dataset.

Figure~\ref{fig:sim_map_point} highlights the difficulty of fitting a spatial model to this
dataset.  Both the ``true'' values, and the perturbed estimates exhibit very different spatial
patterns for the different margins, White children ages 2--3, and American Indian or Alaska
Native children, ages 2--3.  Recall that the largest numbers of White children are in Hennepin
county, which includes Minneapolis, and the largest number of American Indian or Alaska Native
children are located in the more rural Beltrami County in the northern part of the state.
However, predictions made using the multivariate spatial mixture model are able to largely
preserve the spatial patterns in the margins of the dataset, in contrast to predictions made
using the multivariate spatial model.  Also, the predicted values are very close to the
corresponding direct estimates when the area-specific sample size is very large (or
equivalently, when the variance of the direct estimate is very small).  For areas with smaller
sample size, the direct estimates lack precision, and the model-based predictions ``borrow
strength'', by exploiting multivariate and spatial dependence in the data, which is an
important property of general small area estimates.

\begin{figure}[t]
  \centering
  \includegraphics[height=3in, width=0.65\textwidth]{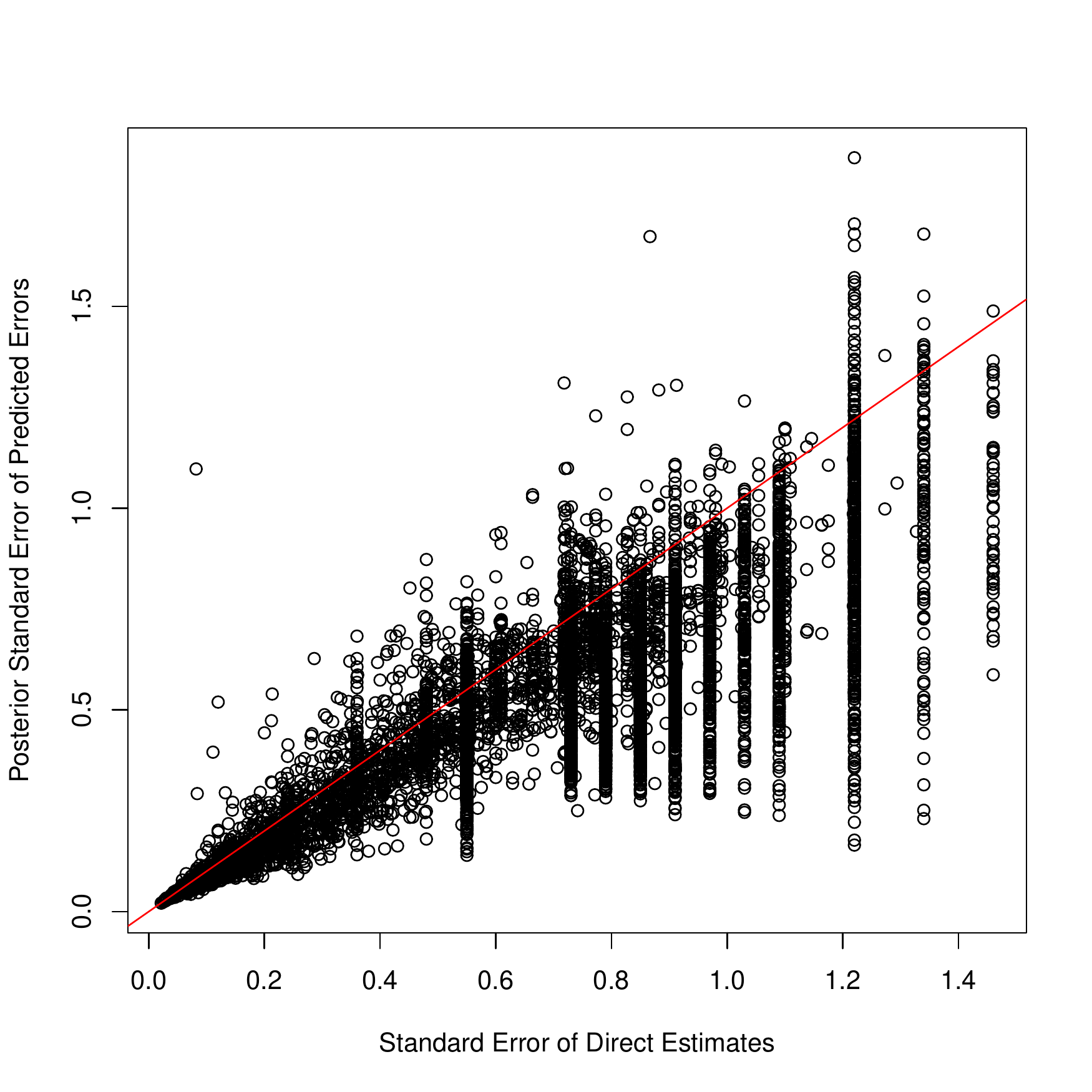}
  \caption{Posterior standard error of predicted values vs. design-based standard
    error of the direct estimates in Simulation 1.}
  \label{fig:sim_sd}
\end{figure}

Figure~\ref{fig:sim_sd} compares the posterior standard error of the predicted values of the
log counts to the design-based standard errors of the pseudo data from the first simulated dataset,
over all counties \(A \in \mathcal{D}\) for all age by race cross-classifications.  From
Figure \ref{fig:sim_sd}, for the majority of the areas, we can see greatly reduced standard
errors of the predicted values compared to the standard errors of the direct estimates, which
is a primary goal.  There are, however, some areas where the standard error of predicted values
increased, and a few of these increases are quite large.

A closer investigation of the predictions with large increases in standard error can be done by
looking at their traceplots and their cluster memberships over the iterations of the Gibbs
sampler.  In these traceplots, it can be seen that these values do not settle into a single
cluster with high probability, but instead switch between multiple clusters.  This has the
effect of limiting the bias of the predictions, but causes increased variability.  In the
analysis of this dataset, this seems to happen relatively infrequently.


Despite these few `outlier' values in Figure~\ref{fig:sim_sd}, it is clear that the MSMM
can effectively cluster the data based on the different spatial patterns and multivariate
characteristics of the data, and can produce predictions which preserve the spatial
characteristics of the data and can exploit spatial and multivariate dependencies so that the
precision of the predictions is greatly increased over the precision of the direct estimates,
for the majority of the areas.

\begin{figure}[t]
  \centering
  \begin{subfigure}{0.45\textwidth}
    \centering
    \includegraphics[height=2in, width=\textwidth]{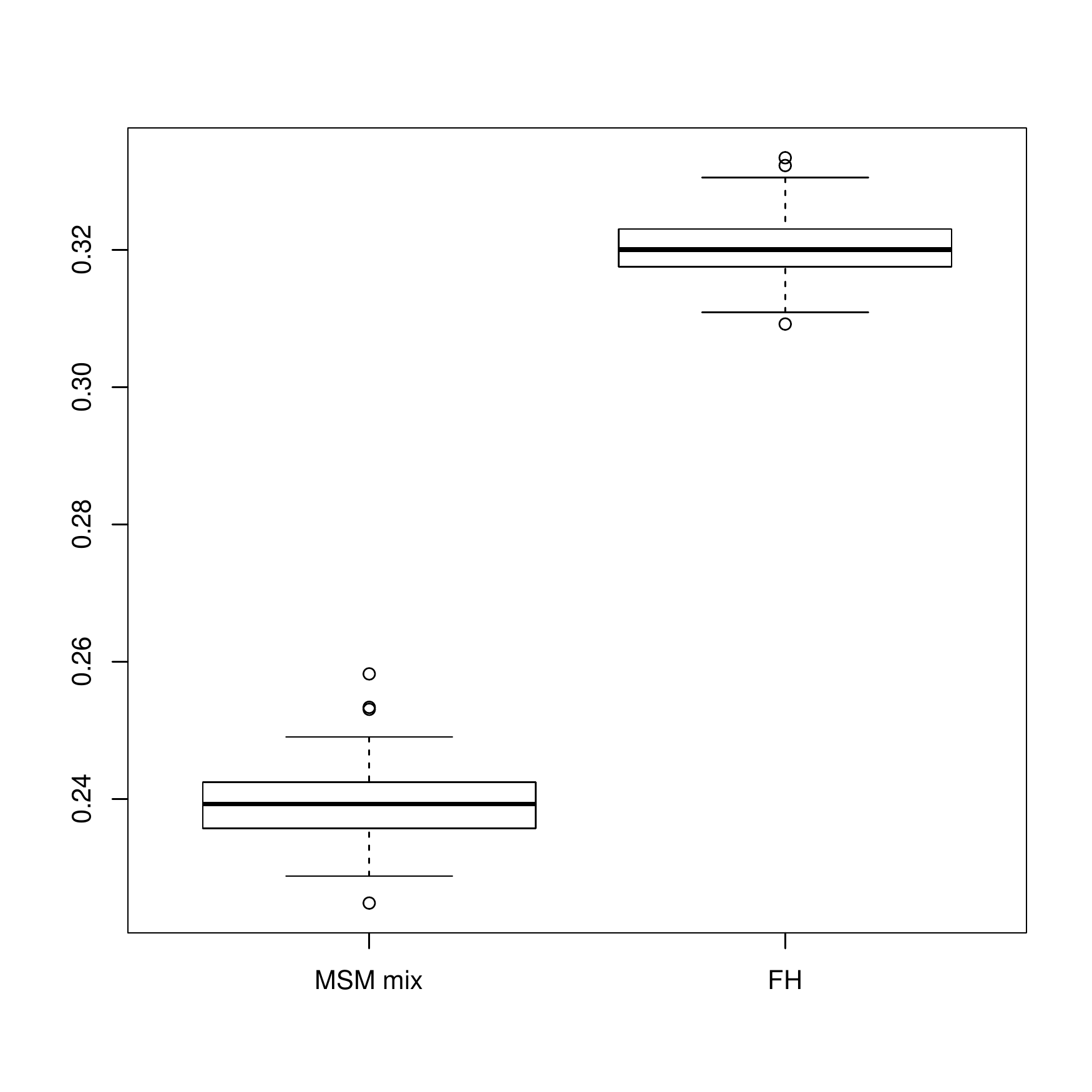}
    \caption{Median absolute bias.}
    \label{fig:simBias}
  \end{subfigure}
  ~
  \begin{subfigure}{0.45\textwidth}
    \centering
    \includegraphics[height=2in, width=\textwidth]{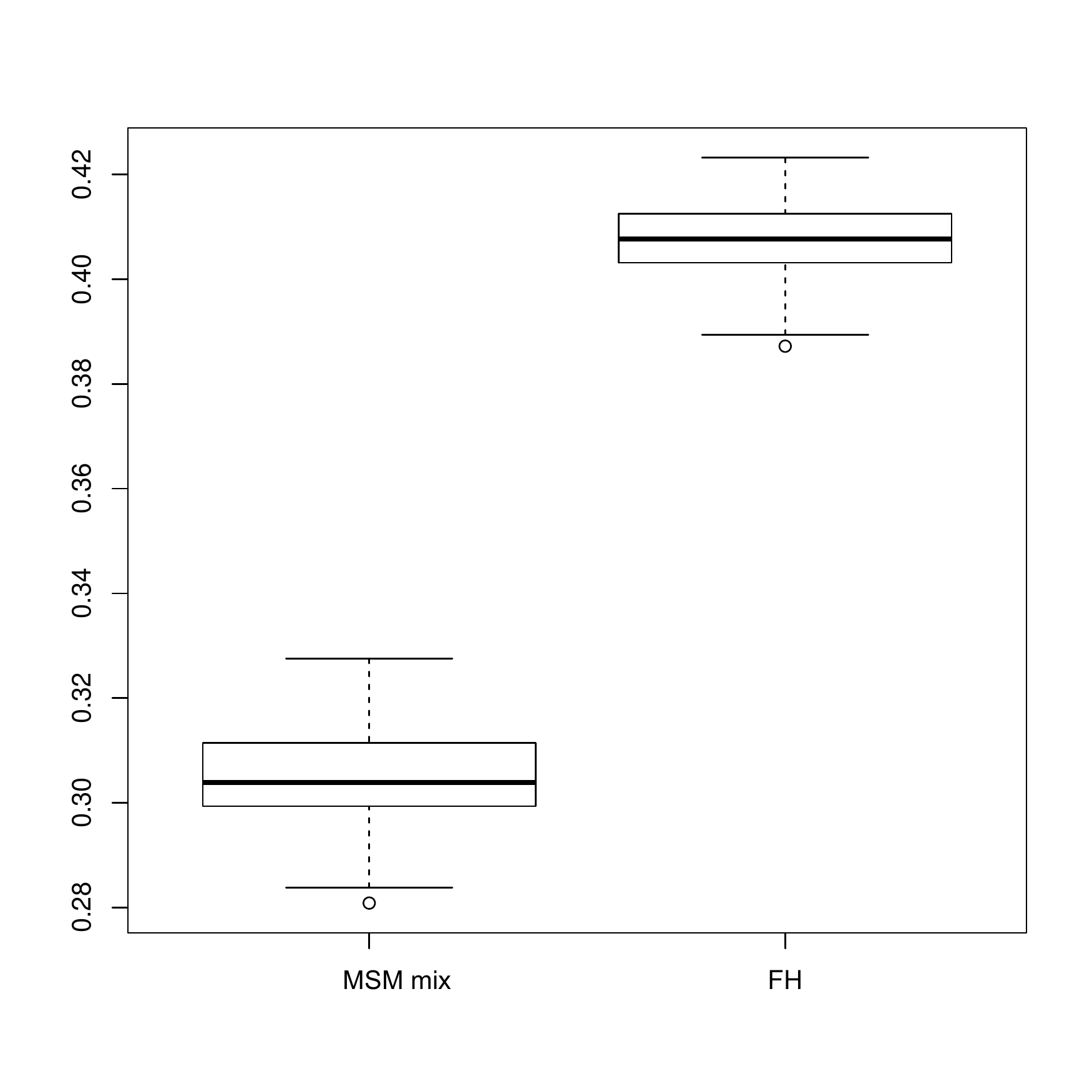}
    \caption{Average mean squared error.}
    \label{fig:simMSE}
  \end{subfigure}
  \caption{Boxplots of the median absolute bias $\text{MAB}_i$ and the average mean squared
    error $\text{AMSE}_i$ for simulation runs $i = 1, \ldots, 100$ from the model-based
    predictions using the MSM model and the Fay-Herriot model.  For comparison, the average
    design-based variance of the direct estimates is 0.54.}
  \label{fig:simResults}
\end{figure}

The overall performance of the predictions using the MSMM can be evaluated by comparing the
predicted values to the true values over the 100 simulations.  First, the median absolute bias,
given by
\begin{equation*}
  \text{MAB}_i = \text{median} \left\{ \left| \hat{Z}_i^{(l)} (A) - Z_i^{(l)} (A)
    \right| : A \in \mathcal{D}, l = 1, \dots, L \right\},
\end{equation*}
is evaluated over the 100 simulations used for predictions using both the MSMM, as well as
predictions using the Fay-Herriot model.  Figure~\ref{fig:simBias} shows a boxplot of the
median absolute bias of predictions from the MSMM compared to the median absolute bias of
predictions from the Fay-Herriot model. From \ref{fig:simBias}, we see that the MSMM
produces predictions with reduced absolute bias compared to predictions from the Fay-Herriot
model.  We also compute the average mean squared error (MSE), given by
\begin{equation*}
  \text{AMSE}_i = \frac{1}{n} \sum_{l=1}^L \sum_{A \in \mathcal{D}} \left(
    \hat{Z}_i^{(l)} (A) - Z_i^{(l)} (A) \right)^2.
\end{equation*}
Figure~\ref{fig:simMSE} shows a boxplot of the average MSE of predictions from the MSMM
compared to the average MSE of predictions from the Fay-Herriot model.  For comparison, the
average design-based variance of the direct estimates is 0.54.  While predictions made using
Fay-Herriot model are, on average, more precise that the direct estimates, the gain in
precision is modest.  It is clear from Figure \ref{fig:simMSE}, that the multivariate and
spatial dependence in the data can be used to greatly improve the MSE of predicted values.

\section{Estimates of the number of children by race and ethnicity in counties in
  Minnesota using the multivariate spatial mixture model}
\label{sec:s48405}

In Section~\ref{sec:msmfail}, we fit the multivariate spatial model to the age by race dataset,
and found that many model-based predictions were unreasonable.  We hypothesized that the cause
of these unrealistic predictions was the presence of multiple spatial fields within the margins
of the multivariate data.  In this situation, it seems plausible that when a model with a
single spatial field is specified, and estimated from the data, that predictions based on the
single spatial field can be too drastically distorted from the direct estimates, when in truth
there are multiple underlying spatial patterns.

In this section, we fit the MSMM to the ACS dataset considered in Section~\ref{sec:msmfail} to
obtain model-based predictions of counts by age and race in counties in Minnesota.  A partial
table of the data is shown in Table~\ref{tab:s48405}.  The data is presented using FIPS codes
for the 376 counties within Minnesota and the surrounding Midwestern states (North Dakota,
South Dakota, Wisconsin, and Iowa).

\begin{table}[t]
  \caption{A selection of the direct estimates and their estimated standard errors of the
    counts of persons by age and race in counties in the Midwestern states surrounding
    Minnesota.  The Order column indexes the 21 age by race categories.  The full dataset
    consists of 7896 rows.}
  \label{tab:s48405}
  \centering
  \begin{tabular}{@{}ccccc@{}}
    \toprule
    State & County & Order & Count & Std.~Err. \\
    \midrule
    19    & 041    & 1     & 325   & 49.2      \\
    19    & 041    & 2     & 370   & 48.0      \\
    19    & 041    & 3     & 375   & 49.9      \\
    \vdots   & \vdots    & \vdots   & \vdots   & \vdots       \\
    46    & 123    & 19    & 4     & 3.6       \\
    46    & 123    & 20    & 15    & 14.0      \\
    46    & 123    & 21    & 15    & 14.0      \\
    \bottomrule
  \end{tabular}
\end{table}

Figure~\ref{fig:s48405both} shows the spatial patterns of the direct estimates within counties
in Minnesota for two margins of the data.  Figure~\ref{fig:s48405direct1} shows the direct
estimates of the number of White children, ages 2--3, and Figure~\ref{fig:s48405direct2} shows
the direct estimates of the number of American Indian or Alaska Native children, ages 2--3.
Exploratory analysis of each of the margins of the data indicate strong spatial correlation
using Moran's I statistic \citep{mor50}.  However, clearly the spatial patterns across
different margins can be quite different.  The different ranges of the data within the
different margins (0--700 in Figure~\ref{fig:s48405direct1} and 0--16,000 in
Figure~\ref{fig:s48405direct2}) pose an additional challenge.  As was seen in
Section~\ref{sec:msmfail}, using a multivariate spatial model with a common spatial field for
all margins of the data for difficult datasets such as this one can result in predictions which
are nonsensical for certain areas, due to the aggressive smoothing from the common fitted
spatial field.

To account for the varying spatial patterns in this dataset, we fit the MSMM to the log of the
direct estimates, plus 1, as in Equation \eqref{E:logDirect}.  This allows for clustering of
the data by common spatial and multivariate characteristics.  It also makes it possible to
obtain model-based predictions which are improvements over the corresponding direct estimates,
by exploiting the multivariate dependence and the spatial dependence within the data.  This is
of particular importance in the situation where there is limited covariate information which
can be used for prediction.

The design-based variance of the log transformed direct estimates was estimated using the
method of replicate weights \citep{jud90}, when these quantities were well defined.  For the
remaining areas, the variance estimates were imputed, using predictions from a LOESS
regression, as was done in Section~\ref{sec:s48400}.  These variance estimates (not shown, as
these are not publicly available data) are used in the data model, and are treated as known
quantities.  The covariates, basis functions, and hyperparameters used were the same as
described in Section~\ref{sec:empirical}.

The MCMC algorithm was run for 10,000 iterations, with the first 5,000 iterations discarded.
As the parameters of interest are the finite population totals, exponential transformation
\eqref{E:inverse} is applied to the model-based predictions (which have been computed at the
log scale) for each iteration of the MCMC chain. Two parallel MCMC chains were run, which
allows for convergence checks on the
parameters \(\alpha\) and \(\sigma^2_{\boldsymbol{\eta}}\), which are not dependent on cluster
labels.  There was no lack of convergence detected based on visual inspection of the MCMC
chains for these parameters, nor was there any indication of convergence issues based on the
batch means, Geweke statistic \citep{gew92}, or the Gelman Rubin statistics \citep{gel92}.  The
computational time to run 10,000 MCMC iterations using a single processor and 25GB of RAM on a
linux server was approximately 14 hours.

All predictions presented here are of aggregated quantities at the county level, which are
invariant to cluster permutations, so we were able to present results without concern about any
possible issues due to label switching.  However, if inference on marginal posterior
quantities, such as cluster membership, is of interest, it is crucial to post-process the
estimates using an algorithm to correct for label switching, for example using the results of
\citet{ste00}.

\begin{figure}[t]
  \centering
  \begin{subfigure}{0.45\textwidth}
    \centering
    \includegraphics[height=2in,width=\textwidth]{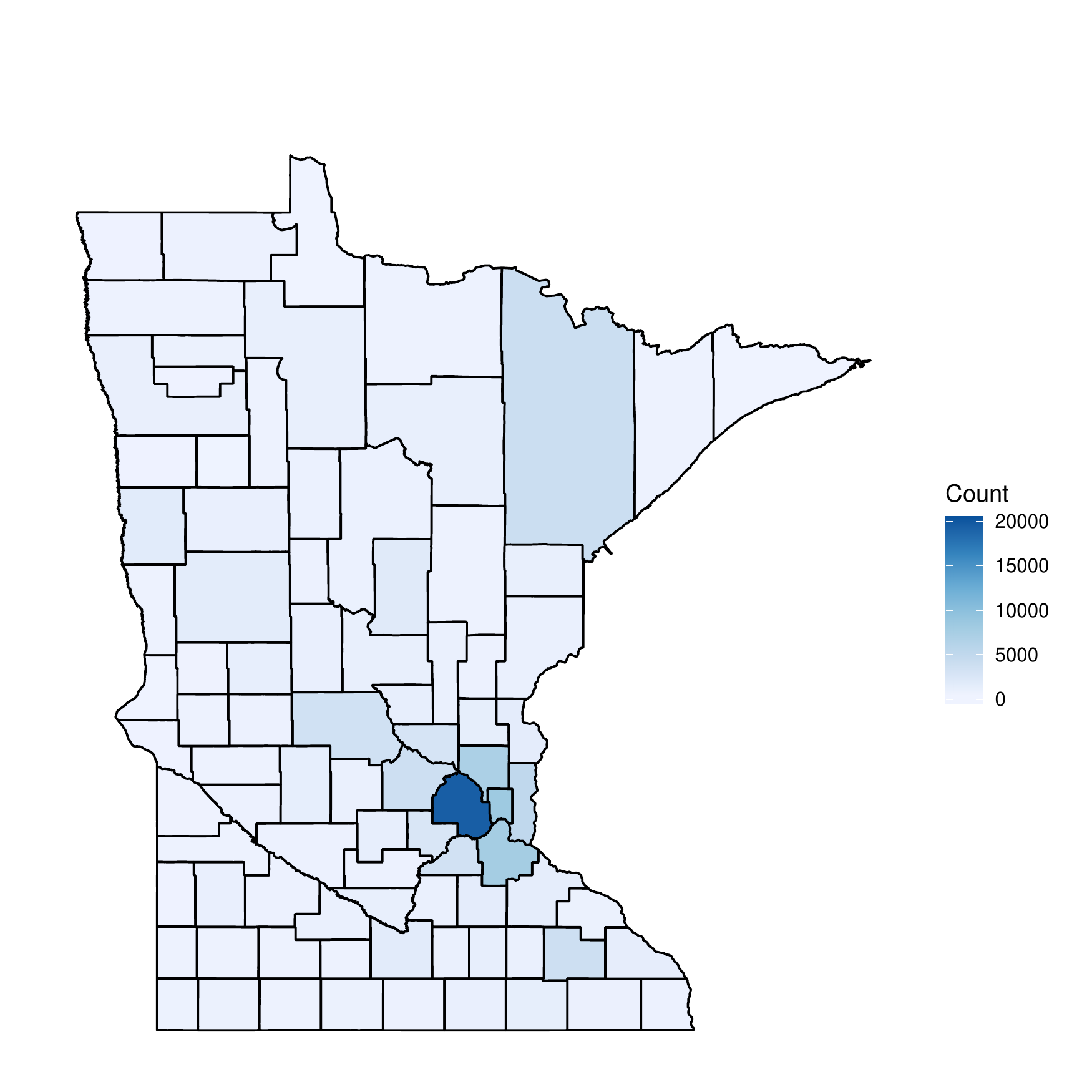}
    \caption{Direct estimates of the number of White children, ages 2--3.}
    \label{fig:s48405direct3}
  \end{subfigure}
  ~
  \begin{subfigure}{0.45\textwidth}
    \centering
    \includegraphics[height=2in,width=\textwidth]{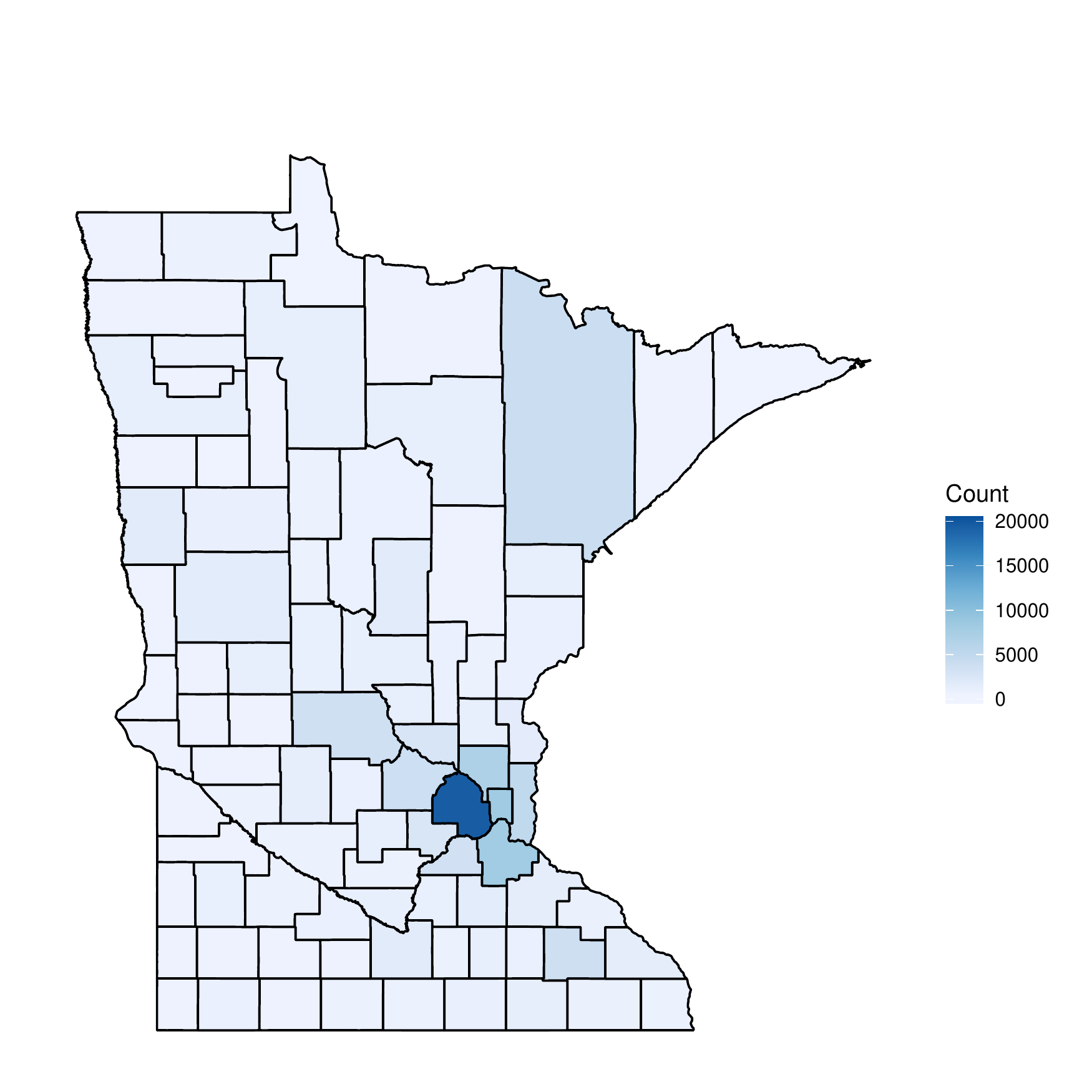} 
    \caption{Model-based predictions of the number of White children, ages 2--3.}
    \label{fig:s48405pred3}
  \end{subfigure}

  \begin{subfigure}{0.45\textwidth}
    \centering
    \includegraphics[height=2in, width=\textwidth]{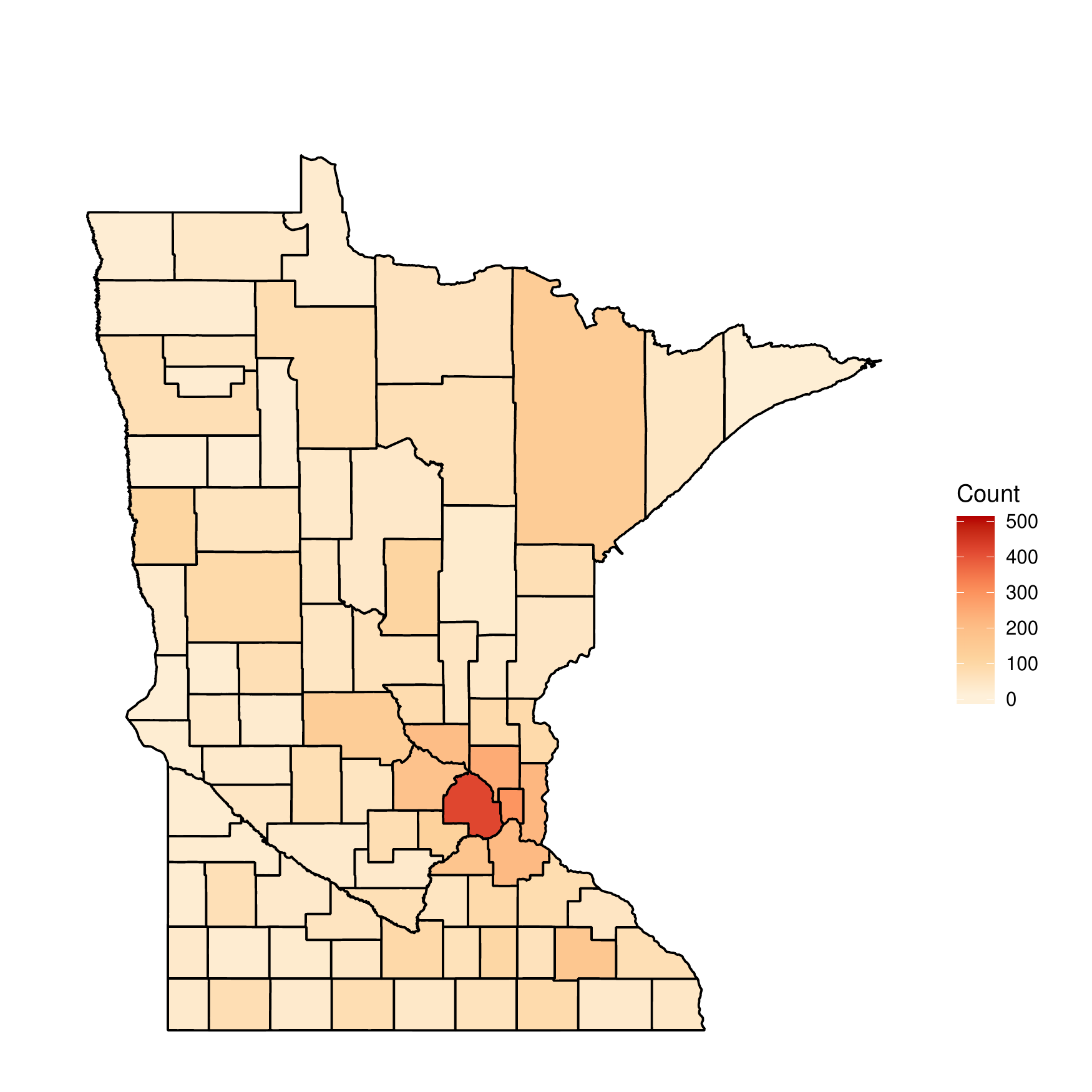}
    \caption{Standard errors of direct estimates of the number of White children,
      ages 2--3.}
    \label{fig:s48405direct3_se}
  \end{subfigure}
  ~
  \begin{subfigure}{0.45\textwidth}
    \centering
    \includegraphics[height=2in, width=\textwidth]{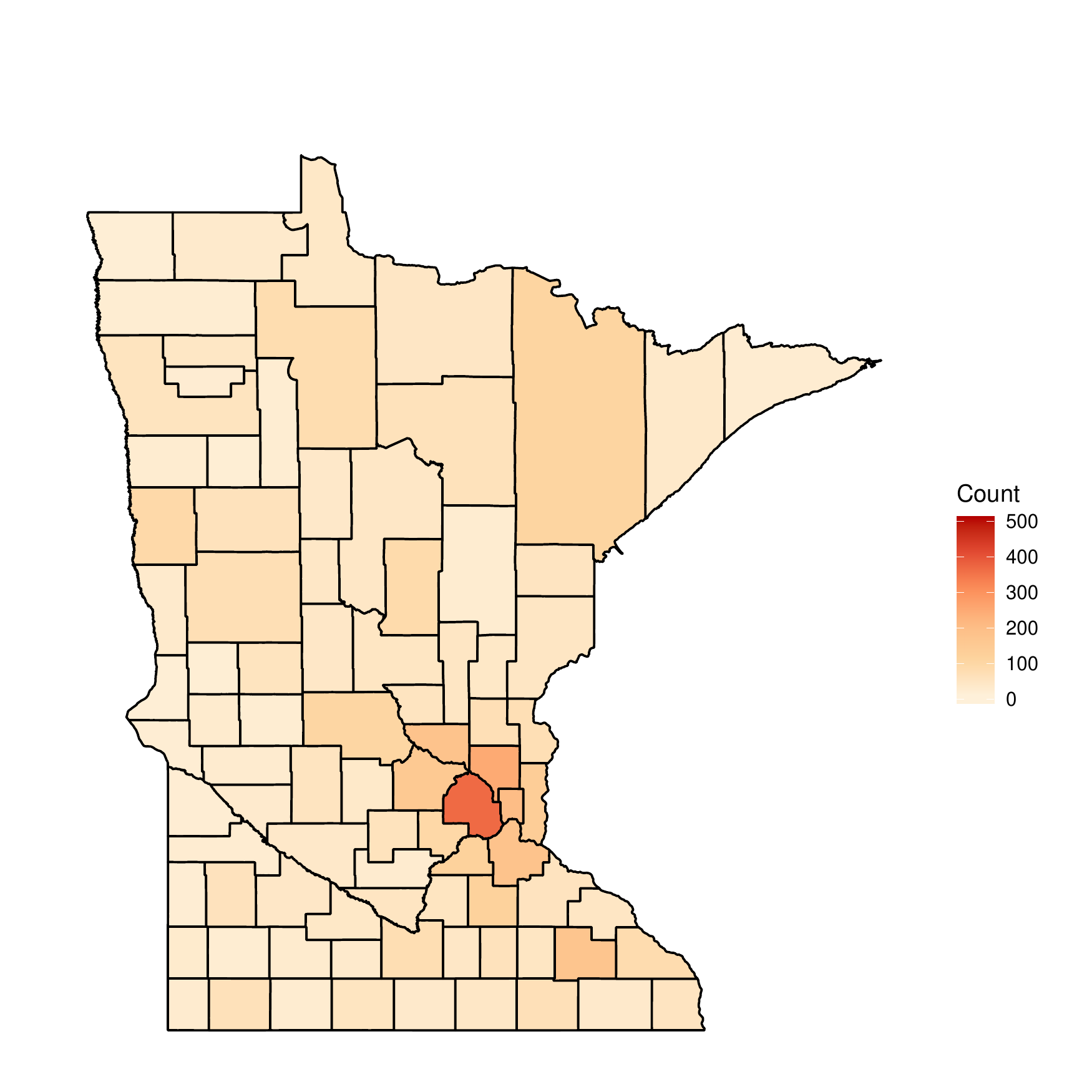}
    \caption{Posterior standard error of model-based predictions of the number of
      White children, ages 2--3.}
    \label{fig:s48405pred3_se}
  \end{subfigure}
  \caption{Comparison of the spatial patterns of the direct estimates and the predicted values
    of the number of White children, ages 2--3, in blue, and a comparison of the spatial
    patterns of the standard errors of the direct estimates and predicted values, of the number
    of White children, ages 2--3, in red.}
  \label{fig:s48405all3}
\end{figure}

\begin{figure}[t]
  \centering
  \begin{subfigure}{0.45\textwidth}
    \centering
    \includegraphics[height=2in, width=\textwidth]{figures/S48405_12_MN_DIRECT}
    \caption{Direct estimates of the number of American Indian or Alaska Native
      children, ages 2--3.}
    \label{fig:s48405direct12}
  \end{subfigure}
  ~
  \begin{subfigure}{0.45\textwidth}
    \centering
    \includegraphics[height=2in, width=\textwidth]{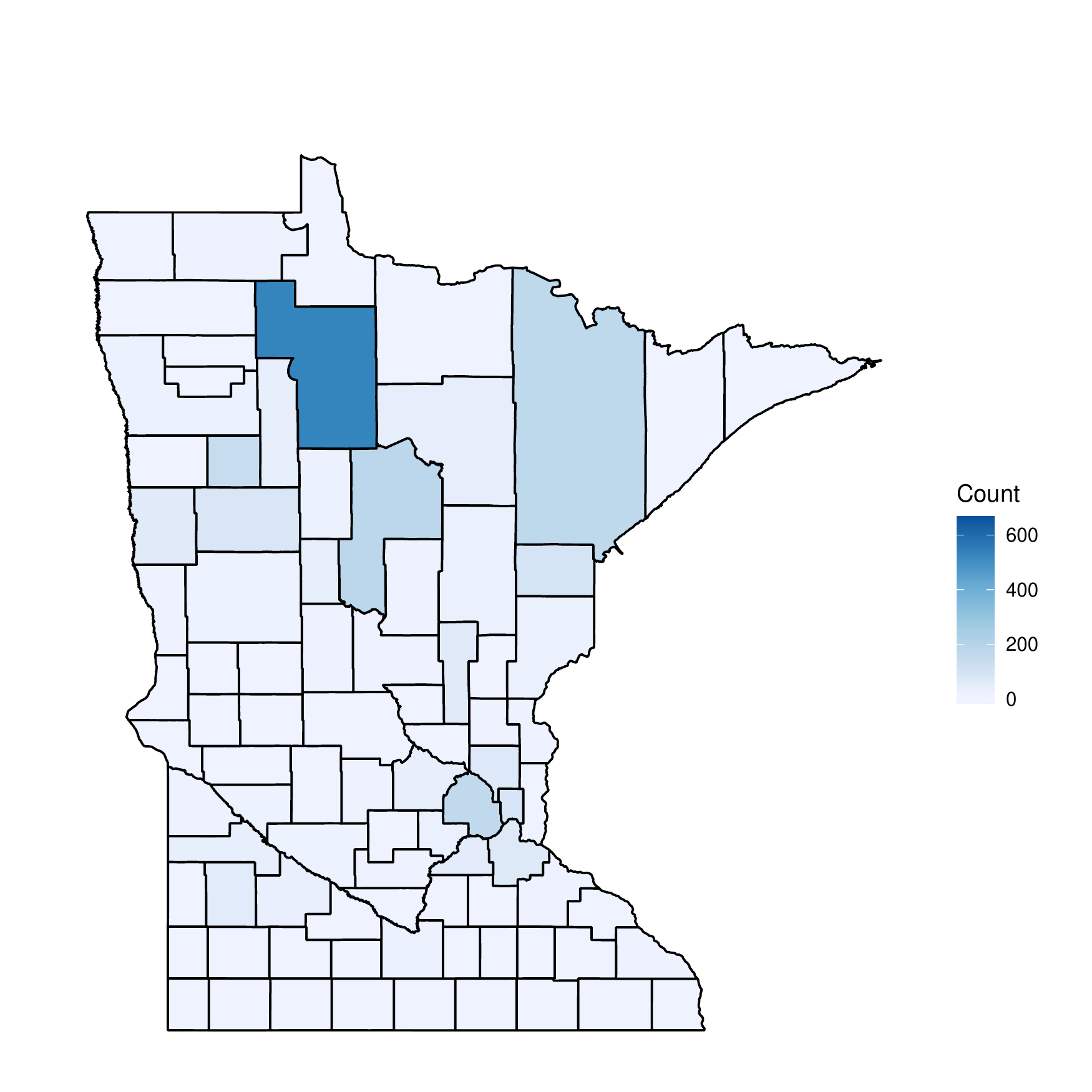}
    \caption{Model-based predictions of the number of American Indian or Alaska
      Native children, ages 2--3.}
    \label{fig:s48405pred12}
  \end{subfigure}

  \begin{subfigure}{0.45\textwidth}
    \centering
    \includegraphics[height=2in, width=\textwidth]{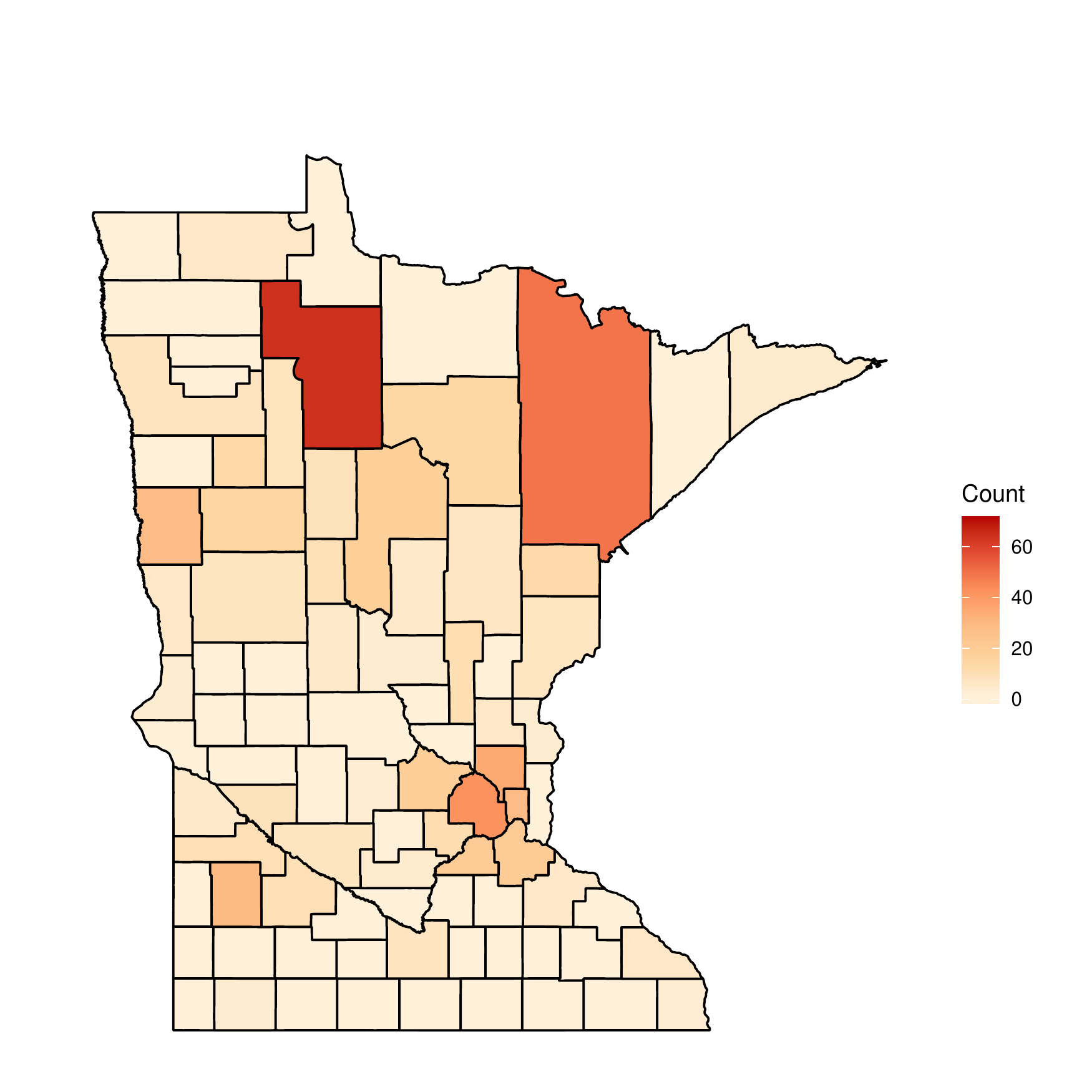}
    \caption{Standard errors of direct estimates of the number of American Indian or
      Alaska Native children, ages 2--3.}
    \label{fig:s48405direct12_se}
  \end{subfigure}
  ~
  \begin{subfigure}{0.45\textwidth}
    \centering
    \includegraphics[height=2in, width=\textwidth]{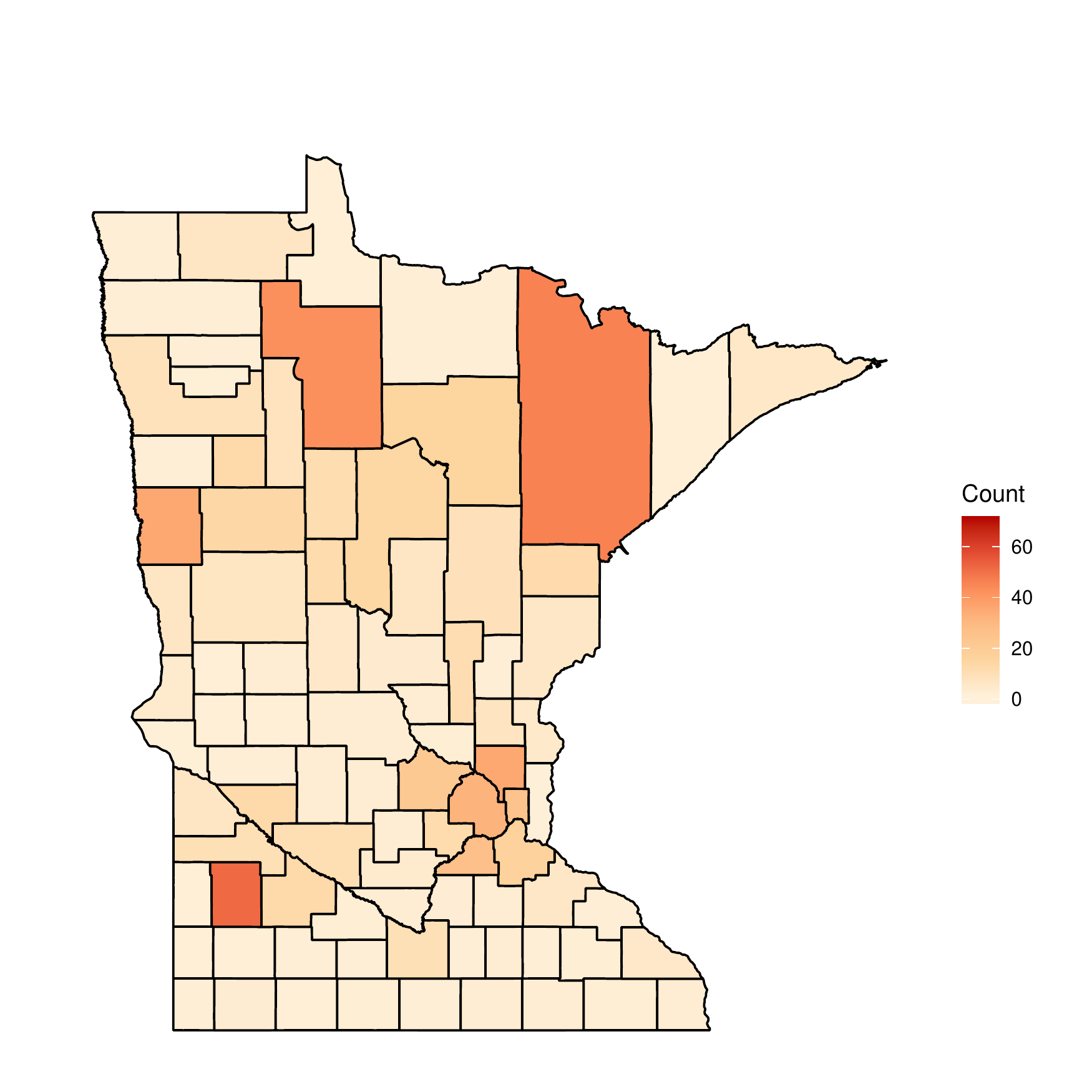}
    \caption{Posterior standard error of model-based predictions of the number of
      American Indian or Alaska Native children, ages 2--3.}
    \label{fig:s48405pred12_se}
  \end{subfigure}
  \caption{Comparison of the spatial patterns of the direct estimates and the predicted values
    of the number of American Indian or Alaska Native children, ages 2--3, in blue, and a
    comparison of the spatial patterns of the standard errors of the direct estimates and
    predicted values, of the number of American Indian or Alaska Native children, ages 2--3, in
    red.}
  \label{fig:s48405all12}
\end{figure}

The nonparametric Dirichlet prior was incorporated into the multivariate spatial mixture model,
in part, to allow the `true' number of clusters to be an unknown quantity.  We note that the
posterior modal number of clusters is 8, with a posterior standard error of 1.  Recall that
there are 21 (3 age and 7 race) observations per county.  The effectiveness of the clustering
mechanism can be seen in Figures~\ref{fig:s48405all3} and \ref{fig:s48405all12}.
Figure~\ref{fig:s48405all3} gives a comparison of the direct estimates and their associated
variance estimates with the model-based predictions of the county-level counts of White
children, ages 2 -- 3, and their posterior variances.  Here, we see similar spatial patterns of
both the direct estimates and the model-based predictions.  We also see the overall reduction
in standard errors in Figures~\ref{fig:s48405direct3_se} and \ref{fig:s48405pred3_se}.

Figure~\ref{fig:s48405all12} gives a comparison of the direct estimates and their associated
variance estimates with the model-based predictions of the county-level counts of American
Indian or Alaska Native children, ages 2--3, and their posterior variances.  A comparison of
Figure~\ref{fig:s48405all12} with Figure~\ref{fig:s48405all3} highlights the ability to
preserve multiple spatial patterns within a given dataset when using predictions from the
multivariate spatial mixture model.  A comparison of the standard errors of the direct
estimates with the posterior standard errors of the model-based predictions is made in
Figures~\ref{fig:s48405direct12_se} and \ref{fig:s48405pred12_se}.  For the majority of the
counties in Minnesota, the posterior standard errors of the model-based predictions were lower
than the corresponding standard errors of the direct estimates.  However, for a few of the
counties, there was a slight increase in the estimated standard error.

\begin{figure}[t]
  \centering
  \includegraphics[height=3in, width=0.65\textwidth]{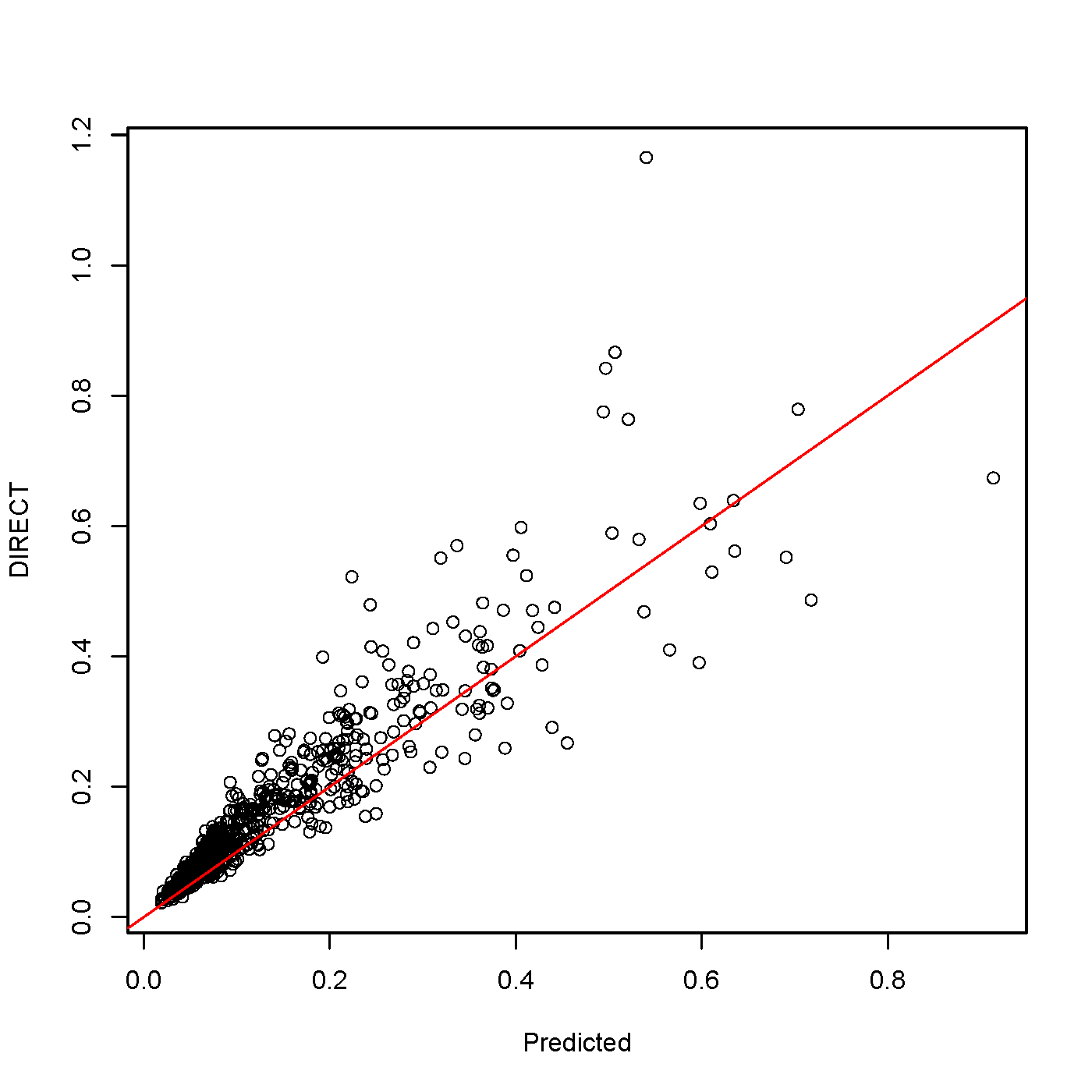}
  \caption{Comparison of the coefficients of variation of the predicted values
    vs. the coefficients of variation of the direct estimates.  The diagonal line is shown in red.}
  \label{fig:s48405cv}
\end{figure}

The overall performance of the uncertainty estimates is presented in Figure~\ref{fig:s48405cv},
which shows a scatter plot of the coefficients of variation of the direct estimates against the
coefficients of variation of the predicted values among the set of nonzero direct estimates.
Figure~\ref{fig:s48405cv} gives a similar result to what was seen in the empirical simulation
study in Section~\ref{sec:empirical}.  Overall, the precision of the model-based predictions is
greater than that of the direct estimates.  However, there are some counties which had an
increase in their coefficient of variation, due to the parameter values associated with that
county switching between clusters at different iterations of the MCMC chain.  This cluster
jumping seems to reduce the bias of the predictions, but can increase variability for some
areas.

While we do see an overall average reduction in the coefficients of variation and posterior
standard errors of the model-based predictions over those corresponding to the direct
estimates, the reductions are not as dramatic as was seen when fitting the multivariate spatial
model to the simpler dataset analyzed in Section~\ref{sec:s48400}.  Also, we do not achieve
uniform reductions in coefficients of variation or standard errors, as there can be increases
in these quantities in individual counties.  Neither of these observations is surprising, as
the age by race by county ACS dataset is of higher dimension and less well behaved than the age
by county dataset.  Likewise, the multivariate spatial mixture model needed to analyze the age
by race dataset is far more complex than the multivariate spatial model.  Despite these
challenges, the multivariate spatial mixture model is promising as a tool for analyzing high
dimensional survey data with varying spatial and multivariate characteristics.  This model
seemed to effectively cluster the age by race dataset in such a way as to preserve spatial
patterns in the data, as well as to produce more precise predictions, on average, than the
corresponding direct estimates.

\section{Conclusion}
\label{sec:conc}
Model-based estimation of area-level tabulations from the ACS is a challenging
problem at the sub-state level when simultaneously considering other detailed demographic
factors. Motivated by the need to produce special tabulations for the ACS, this paper
introduces a nonparametric Bayesian multivariate spatial mixed effects model. As demonstrated
in Sections~\ref{sec:empirical} and \ref{sec:s48405}, the proposed model provides a solution to
an important problem encountered at the U.S. Census Bureau and is of independent interest.

When disseminating model-based estimates (model-based special tabulations) it is critical that
the estimates retain the spatial patterns found in the observed sample and significantly
improve the precision over the direct estimates. Additionally, the model needs to be flexible,
in that it is effective across a wide range of problems without requiring subject matter
expertise to propose new model variants for each dataset considered. These aspects are achieved
by our proposed approach.

Notably, our model is nonparametric Bayes. For extremely high-dimensional settings, we consider
an approximation that uses a finite representation through a stick-breaking prior. The latter
approach may be preferable in a production setting at federal statistical agencies, as this can
improve the computational efficiency without significant loss in precision.

The approach considered here is conducted at the area level. Nevertheless, official statistical
agencies have access to the underlying confidential micro-data (unit-level). One area of future
research is to extend this approach to the unit-level. To do this in the context of ACS would
require methodology that accommodates the informative sampling mechanism.

\if1\blind
{
\begin{center}
{\large\bf Acknowledgements}
\end{center}
The DRB approval number for this paper is CDBRB-FY20-044.  This report is released to
inform interested parties of ongoing research and to encourage discussion of work in
progress.  The views expressed are those of the authors, and not those of
the U.S. Census Bureau.
}\fi

\nocite{*}


\appendix

\end{document}